\documentclass[12pt]{article}
\usepackage{mathtools, amssymb, amsthm, graphicx, float, bm, bbm, threeparttable}
\usepackage[colorlinks = true, allcolors = blue]{hyperref}
\usepackage[round]{natbib}

\newcommand{\blind}{0}
\def\spacingset#1{\renewcommand{\baselinestretch}{#1}\small\normalsize}

\def\argmin{\mathop{\rm argmin}}
\def\argmax{\mathop{\rm argmax}}

\newcommand{\mA}{{\mathcal A}}
\newcommand{\mX}{{\mathcal X}}
\newcommand{\mD}{{\mathcal D}}
\newcommand{\mF}{{\mathcal F}}
\newcommand{\mH}{{\mathcal H}}
\newcommand{\mG}{{\mathcal G}}

\newcommand{\mO}{{\mathcal O}}

\newcommand{\bX}{{\bm X}}
\newcommand{\bx}{{\bm x}}
\newcommand{\by}{{\bm y}}

\newcommand{\bW}{{\bm W}}
\newcommand{\be}{{\bm e}}
\newcommand{\bdf}{{\bm f}}

\newcommand{\bK}{{\bm K}}
\newcommand{\bb}{{\bm b}}
\newcommand{\balpha}{{\bm\alpha}}
\newcommand{\bbeta}{{\bm\beta}}
\newcommand{\bgamma}{{\bm\gamma}}
\newcommand{\bdelta}{{\bm\delta}}
\newcommand{\bSigma}{{\bm\Sigma}}
\newcommand{\bzero}{{\bm0}}
\newcommand{\bone}{{\bm1}}

\newcommand{\bbx}{{\mathbf x}}
\newcommand{\bbK}{{\mathbf K}}
\newcommand{\bbX}{{\mathbf X}}
\newcommand{\bbA}{{\mathbf A}}
\newcommand{\bbP}{{\mathbf P}}
\newcommand{\bbV}{{\mathbf V}}
\newcommand{\bbM}{{\mathbf M}}

\newcommand{\bbB}{{\mathbf B}}
\newcommand{\bbC}{{\mathbf C}}
\newcommand{\bbI}{{\mathbf I}}
\newcommand{\bbJ}{{\mathbf J}}

\newcommand{\R}{{\mathbb R}}
\newcommand{\E}{{\mathbb E}}
\newcommand{\V}{{\mathrm{Var}}}
\newcommand{\bias}{{\mathrm{bias}}}

\newcommand{\diag}{\mathrm{diag}}

\newcommand{\PE}{\mathrm{PE}}
\newcommand{\ind}[1]{\mathbbm{1}{\left[{#1}\right] }}

\newcommand{\<}[1]{\langle{#1}\rangle}
\newcommand{\ie}{i.e.}

\newtheorem{assumption}{Assumption}

\newtheorem{theorem}{Theorem}
\newtheorem{corollary}{Corollary}

\newtheorem{remark}{Remark}

\theoremstyle{remark}
\date{}

\begin{document}

% ----------------------- Cover -------------------------
\spacingset{1}

\if0\blind
{
  \title{\bf Doubly Robust Direct Learning for Estimating Conditional Average Treatment Effect}
  \author{Haomiao Meng and Xingye Qiao\thanks{
    Correspondence to: Xingye Qiao (e-mail: qiao@math.binghamton.edu). Haomiao Meng is a PhD student in the Department of Mathematical Sciences at Binghamton University, State University of New York, Binghamton, New York, 13902-6000.
    Xingye Qiao is an Associate Professor in the Department of Mathematical Sciences at Binghamton University, State University of New York, Binghamton, New York, 13902-6000.}}
  \maketitle
} \fi

\if1\blind
{
  \bigskip
  \bigskip
  \bigskip
  \begin{center}
    {\LARGE\bf Doubly Robust Direct Learning for Estimating Conditional Average Treatment Effect}
  \end{center}
  \medskip
} \fi

\bigskip
\begin{abstract}
\noindent
Inferring the heterogeneous treatment effect is a fundamental problem in the sciences and commercial applications. In this paper, we focus on estimating Conditional Average Treatment Effect (CATE), that is, the difference in the conditional mean outcome between treatments given covariates. Traditionally, Q-Learning based approaches rely on the estimation of conditional mean outcome given treatment and covariates. However, they are subject to misspecification of the main effect model. Recently, simple and flexible one-step methods to directly learn (D-Learning) the CATE without model specifications have been proposed. However, these methods are not robust against misspecification of the propensity score model. We propose a new framework for CATE estimation, robust direct learning (RD-Learning), leading to doubly robust estimators of the treatment effect. The consistency for our CATE estimator is guaranteed if either the main effect model or the propensity score model is correctly specified. The framework can be used in both the binary and the multi-arm settings and is general enough to allow different function spaces and incorporate different generic learning algorithms. As a by-product, we develop a competitive statistical inference tool for the treatment effect, assuming the propensity score is known. We provide theoretical insights to the proposed method using risk bounds under both linear and non-linear settings. The effectiveness of our proposed method is demonstrated by simulation studies and a real data example about an AIDS Clinical Trials study.
\end{abstract}

\vfill

% ----------------------- Main Body -------------------------
\newpage

\spacingset{1.45} % DON'T change the spacing!
\setcounter{page}{1}
\abovedisplayskip = 8pt
\belowdisplayskip = 8pt

\section{Introduction}

Inferring the heterogeneous treatment effect  is a fundamental problem in the sciences and commercial applications. Examples include studies on the effect of certain advertising or marketing efforts on consumer behavior \citep{bottou2013counterfactual}, research on the effectiveness of public policies \citep{turney2015detrimental}, and ``A/B tests'' in the context of tech companies for product development \citep{taddy2016nonparametric}. In particular, it can be useful in personalized medicine: based on many biomarkers, how can we determine which patients can potentially benefit from a treatment \citep{royston2008interactions}?

Under the potential outcome framework \citep{rubin1974estimating,imbens2015causal}, we are interested in the comparison between the observed outcome and the counterfactual outcome we would have observed under a different regime or treatment. Assuming that there are two treatment arms ($\{+1, -1\}$), we want to estimate the difference in the conditional mean outcome between the two treatments, given the individual's pre-treatment covariates. This problem is typically known as Conditional Average Treatment Effect (CATE) estimation. CATE is closely associated with the optimal individualized treatment rule (ITR). The latter maximizes the mean of a (clinical) outcome in a population of interest.

Traditional approaches to developing optimal ITRs, such as the Q-Learning (``Q'' denoting ``quality'') \citep{watkins1992q,  qian2011performance, moodie2014q}, models the conditional mean outcome given the treatment and covariates for each treatment separately, and then takes their difference to estimate the CATE. Note the conditional mean outcome includes the main effect of the treatment and the contrast between the two treatment arms. Q-Learning often requires correct specifications for both the main effect and the contrast, even though only the contrast part really matters with respect to the CATE and ITR estimation. In this case, the main effect is a nuisance parameter, whose specification may affect the estimation of treatment contrast. More robust approaches than Q-Learning have been proposed \citep{lu2013variable} under the A-Learning framework (``A'' denoting ``advantage'') for optimal dynamic treatment regimes \citep{murphy2003optimal,robins2004optimal}. These approaches are robust against the misspecification of the main effect model \citep{schulte2014q}.

Recently, there is a growing literature on using machine learning for CATE or ITR estimation, including regression trees \citep{athey2016recursive,su2009subgroup}, random forests \citep{wager2018estimation}, boosting \citep{powers2018some}, neural nets \citep{johansson2016learning}, Bayesian machine learning  \citep{chipman2010bart, hill2011bayesian,taddy2016nonparametric, hahn2020bayesian}, and combinations of the above \citep{kunzel2019metalearners}. Besides those under the Q-Learning framework, many of these methods can be categorized to \textit{modified outcome methods} and \textit{modified covariate methods}, following the categorization of \citet{knaus2018machine}.  \textit{Modified outcome methods} were proposed in the Ph.D. thesis of James Signorovitch, Harvard University \citep{signorovitch2007identifying} in the randomized experimental setting. In the observational data setting, the inverse probability weighted (IPW) estimator can be used to modify the outcome. Similar modified outcome approaches have been proposed which allows directly using off-the-shelf machine learning algorithms for CATE or ITR estimation \citep{beygelzimer2009offset,dudik2011doubly,weisberg2015post}. A drawback of the IPW estimator is that its performance hinges upon accurate estimation of the propensity score. The doubly robust (DR) augmented inverse probability weighted (AIPW) approach \citep{robins1994estimation,bang2005doubly} was formulated by \citet{zhang2012robust}. It requires an estimation of the treatment propensity score and the conditional mean outcome given treatment and covariates. The double robustness of this method is well studied: as long as the model for either the conditional mean outcomes or the propensity score is correctly specified, the estimator is consistent. See more work on double robustness in \cite{kang2007demystifying, cao2009improving, zhang2013robust, zhao2014doubly, fan2016improving,zhao2019efficient}. Many of these methods focus on estimating optimal ITRs instead of CATE.

The \textit{modified covariate method} was introduced by \citet{tian2014simple} for the experimental setting and was later generalized to the observational setting by \citet{chen2017general}. \citet{qi2018d} proposed a variant of the modified covariate method for estimating the optimal ITR. Modified covariate methods do not require to specify any model of the main effect or the conditional mean outcome function and they directly estimate the CATE. We refer to them as the D-Learning (``D'' for ``direct''). While D-Learning models the treatment effect directly, it relies on an accurate estimate of the propensity score in the observational setting. \citet{tian2014simple} and \citet{chen2017general} both described the possibility to increase the 
efficiency of their estimators. This efficiency augmentation variant replaces the outcome by the residual of the outcome less the conditional mean outcome function. Though such efficiency augmentation has been shown to work well in certain scenarios, a double robustness property has not yet been discovered among \textit{modified covariate methods}. We are not aware of any further theoretical analyses
of the statistical properties of this approach.

The first contribution of this paper is a modified-covariate (D-Learning)  method with a double robustness property for CATE estimation. Compared to the aforementioned efficiency augmentation using the \textit{conditional mean outcome function}, we find that one should replace the outcome by the residual with respect to the \textit{main effect function} to achieve the double robustness. 
Different from the double robustness in the AIPW estimator literature (which protects against misspecification of the treatment-specific conditional mean outcomes and the propensity score), our method is robust with respect to the main effect and the propensity score. More precisely speaking, the consistency for our CATE estimation is guaranteed if either the main effect model or the propensity score model is correctly specified. Our method does not require the estimation of the treatment-specific conditional mean outcome. 

The second goal of this paper is to generalize our method to the multi-arm case. \citet{qi2018d,qi2019multi} discussed the D-Learning in the multi-arm case, but their method mainly focused on ITR estimation (instead of CATE) and did not have a robustness property. Thirdly, we provide a theoretical analysis of the convergence rate for the prediction error of our CATE estimation. Moreover, we consider a special setting with known propensity scores, in which case, we propose an efficient estimator for the main effect and an unbiased estimator for the treatment effect, and derive the asymptotic normality which affords statistical inference.

A related result to our work can be found in \citet{nie2017quasi} and \citet{shi2016robust}, which we refer to as the R-Learning, and offer an optimization problem in the form of A-Learning (``R'' refers to Robinson's transformation in \citet{robinson1988root}, ``residual'', and ``robust''). They modify both the outcome and the covariates, and offer some robustness protection. However, neither enjoys the double robustness property. Our method can be viewed as the robustified version of the D-Learning; hence, we call it the RD-Learning.

The rest of the paper is organized as follows. In Section \ref{sec:background}, we introduce some notations and background. We present the proposed RD-Learning method in Section \ref{sec:rd-learn}. Statistical inference in the known propensity score setting can be found in Section \ref{sec:inference}. In Section \ref{sec:simulation}, we design simulation studies to validate the proposed method, followed by a real data example on AIDS clinical trial in Section \ref{sec:real}. Section \ref{sec:conclusion} concludes the paper.

\section{Notations and Background}\label{sec:background}

First consider a two-arm randomized trial. A patient, with pre-treatment covariate $\bX \in \mX \subseteq \R^p$, is randomly assigned to treatment $A \in \mA = \{1, -1\}$. Let $Y^*(j) \in \R$ be the potential outcome the patient would receive by receiving treatment $j \in \mA$. The observed clinical outcome is denoted by $Y = Y^*(A)$. Let $p_j(\bx) = \mathbbm{P}(A = j \mid \bX = \bx)$. Assumption \ref{asp:reg} is a typical regularity assumption.

\begin{assumption}\label{asp:reg}
For any $j \in \mA$, $Y^*(j) \perp A \mid \bX$ and $p_j(\bx) \geq c$ for some $c \in (0, 1)$.
\end{assumption}

Let $P$ be the distribution of the triplet $(\bX, A, Y)$. The goal is to estimate the Conditional Average Treatment Effect (CATE), defined as
$$\E(Y^*(1) - Y^*(-1) \mid \bX = \bx),$$
based on a training sample $\{(\bx_i, a_i, y_i)\}_{i=1}^n$ randomly drawn from $P$.

It is typical to consider the following model,
\begin{align}\label{model_binary}
	Y = m(\bX) + A\delta(\bX) + \epsilon, \quad \text{where}\ \E(\epsilon) = 0, \ \V(\epsilon) = \sigma^2 < \infty.
\end{align}
Denote the conditional mean outcome as $\mu_j(\bx) \triangleq \E(Y^*(j) \mid \bX = \bx) = \E(Y \mid \bX = \bx, A = j)$. It can be easily verified that the main effect $m(\bx) = (\mu_1(\bx) + \mu_{-1}(\bx))/2$, and the treatment effect $\delta(\bx) = (\mu_1(\bx) - \mu_{-1}(\bx))/2$. Thus, to estimate CATE is equivalent to to estimate $\delta(\bx)$. In this article, we refer to $\delta(\bx)$ as the treatment effect.

One way to estimate $\delta(\bx)$ is to conduct regression modeling for $\mu_j(\bx)$, $j \in \{1, -1\}$. This approach is known as Q-Learning \citep{murphy2005generalization, qian2011performance}, where $\mu_j(\bx)$ is referred to as the ``Q function". For example, one may consider linear regression models for $\mu_j(\bx)$ which imply linear modeling for both $m(\bx)$ and $\delta(\bx)$, such as $m(\bx) = \bbx^T\balpha$ and $\delta(\bx) = \bbx^T\bbeta$ with $\bbx = (1, \bx^T)^T\in\R^{p+1}$. The coefficients are estimated by solving the following optimization problem,
$$\min_{\balpha, \bbeta \in \R^{p+1}} \frac{1}{n} \sum_{i=1}^n (y_i - \bbx_i^T\balpha - a_i \bbx_i^T\bbeta)^2.$$
This approach may be vulnerable to model mis-specification of $m(\bx)$ and $\delta(\bx)$. A partial solution is to consider a broader model space (e.g. non-parametric models) to avoid model mis-specification.

\cite{tian2014simple} proposed a new method to estimate $\delta(\bx)$ without specifying the model for $m(\bx)$ under the completely randomized trail setting, \ie, $p_1(\bx) = 1/2$. By observing that $\E(AY \mid \bX = \bx) = \delta(\bx)$, we may use a linear function $\bbx^T\bbeta$ to model $\E(AY \mid \bX = \bx)$ directly. \cite{chen2017general} considered a more general framework to accommodate other proportion $p_1(\bx)$ than $1/2$, as well as observational studies. Specifically, for linear modeling, the treatment effect $\delta(\bx)$ is estimated by $\bbx\hat\bbeta$ where 
\begin{align}\label{d-learn}
	\hat\bbeta = \argmin_{\bbeta \in \R^{p+1}} \frac{1}{n} \sum_{i=1}^n \frac{1}{p_{a_i}(\bx_i)} (a_iy_i - \bbx_i^T\bbeta)^2 = \argmin_{\bbeta \in \R^{p+1}} \frac{1}{n} \sum_{i=1}^n \frac{1}{p_{a_i}(\bx_i)} (y_i - a_i \bbx_i^T\bbeta)^2.
\end{align}
This estimator has been proved to be consistent under Assumption \ref{asp:reg}. Unlike Q-Learning, in which the estimator to $\delta(\bx)$ is based on the estimators for $\mu_j(\bx)$'s, this approach directly estimates the treatment effect $\delta(\bx)$. Hence, it is named Direct Learning or D-Learning \citep{qi2018d}. Non-linear or sparse modeling is also possible in this framework.

One advantage of D-Learning over Q-learning is that it avoids mis-specification of the main effect $m(\bx)$. However, existing consistency results for D-Learning assume that the propensity score $p_j(\bx)$ is known or at least correctly specified, which may not be satisfied in observational studies. Moreover, as will be shown later, the D-Learning estimator also suffers a larger variance compared to other methods.

The AIPW estimator is a well-studied approach that can address the misspecification issue of the propensity score. It was first proposed to estimate the (unconditional) average treatment effect, $\Delta\triangleq \E(Y^*(1) - Y^*(0))$ using $\hat\Delta \triangleq \hat\psi_1 - \hat\psi_{-1}$, where 
$$\hat\psi_j = \frac{1}{n} \sum_{i=1}^n \frac{y_i\ind{a_i = j}}{\hat p_j(\bx_i)} - \frac{\ind{a_i = j} - \hat p_j(\bx_i)}{\hat p_j(\bx_i)} \hat\mu_j(\bx_i)\quad j \in \{1, -1\},$$
where $\hat p_j(\bx)$ is an estimator for $p_j(\bx)$ and $\hat\mu_j(\bx)$ is an estimator for $\mu_j(\bx)$. The AIPW estimator has a double robustness property in that $\hat\Delta$ is consistent if either $p_j(\bx)$ or $\mu_j(\bx)$ is correctly specified for each $j$. The main product of this paper is a D-Learning method for CATE estimation with a similar doubly robust property.

\section{RD-Learning}\label{sec:rd-learn}

We first introduce our proposed Robust Direct Learning (RD-Learning) approach for the binary case, then generalize it to the multi-arm case. This is followed by the theoretical study of the proposed method.

\subsection{RD-Learning in the Binary Case}\label{sec:binary}

Given a training sample $\{\bx_i, a_i, y_i\}_{i=1}^n$, the RD-Learning method is based on an estimator for the propensity score $p_1(\bx)$, denoted by $\hat p_1(\bx)$, and an estimator for the main effect $m(\bx)$, denoted by $\hat m(\bx)$. They can be any existing estimators commonly used in the literature. If we consider linear modeling for the treatment effect, \ie, $\delta(\bx) = \bbx\bbeta$, then RD-Learning estimator for $\bbeta$ is obtained by solving
\begin{align}\label{rd-learn}
	\hat\bbeta = \argmin_{\bbeta \in \R^{p+1}} \frac{1}{n} \sum_{i=1}^n \frac{1}{\hat p_{a_i}(\bx_i)} (y_i - \hat m(\bx_i) - a_i\bbx_i^T\bbeta)^2,
\end{align}
where $\bbx_i \triangleq (1, \bx_i^T)^T$, and the treatment effect is estimated by $\hat\delta(\bx) = \bbx^T\hat\bbeta$.

Comparing (\ref{rd-learn}) with (\ref{d-learn}), the RD-Learning is an augmented version of D-Learning by replacing the outcome $y_i$ in (\ref{d-learn}) with the residual $y_i - \hat m(\bx)$. In the literature, similar procedures in which the outcome is replaced by a certain residual have been proposed to improve the efficiency of the estimation. For example, R-Learning methods \citep{shi2016robust,nie2017quasi} replaced the outcome $y_i$ in the A-Learning framework \citep{murphy2003optimal, robins2004optimal} by the residual $y_i - \hat\Phi(\bx_i)$, where $\hat\Phi(\bx)$ is an estimator for the conditional mean outcome function $\Phi(\bx)\triangleq\E(Y \mid \bX = \bx)$. In the ITR literature, residual weighted learning \citep[RWL]{zhou2017residual} replaced the outcome $y_i$ in the outcome weighted learning \citep[OWL]{zhao2012estimating} by residual $y_i - \hat m(\bx_i)$ with respect to the main effect estimator $\hat m(\bx_i)$. In general, these procedures reduce the variance of the estimator. Moreover, it has been shown that in these works \citep{shi2016robust,nie2017quasi,zhou2017residual}, the CATE estimators are still consistent even when $\hat\Phi(\bx)$ or $\hat m(\bx)$ is mis-specified, so long as the propensity score $p_1(\bx)$ is known or can be consistently estimated. In other words, they are robust against misspecification for $\Phi(\bx)$ or $m(\bx)$.

Although residual based approaches have been studied in the aforementioned work, it has not been thoroughly studied for modified-covariate (D-Learning) methods. Specifically, while it is commonly recognized that replacing the outcome by the residual can improve efficiency \citep{tian2014simple}, it was unclear which residual should be used. In RD-Learning (\ref{rd-learn}), we propose to use the residual with respect to the main effect instead of the conditional mean outcome $m(\bx)$. As will be shown below, the proposed RD-Learning estimator is not only robust to misspecification for $m(\bx)$, but also has an ``double robustness" property similar to that of the AIPW estimator \citep{zhang2012robust}. This is described in the following theorem.

\begin{theorem}\label{thm:rd_binary}
Suppose model (\ref{model_binary}) holds. Let $\tilde p_1(\bx)$ be a working model for the propensity score $p_1(\bx)$ with $0 < \tilde p_1(\bx) < 1$ and $\tilde m(\bx)$ be a working model for the main effect $m(\bx)$. Assume that Assumption \ref{asp:reg} holds. Then we have
$$\delta \in \argmin_{f \in \{\mX \to \R\}} \E\left[\frac{1}{\tilde p_A(\bX)} (Y - \tilde m(\bX) - Af(\bX))^2\right]$$
if either $\tilde p_1(\bx) = p_1(\bx)$ or $\tilde m(\bx) = m(\bx)$ for $\bx \in \mX$ almost surely.
\end{theorem}

Theorem \ref{thm:rd_binary} also holds when, the functions $\tilde p_1(\bx)$ and $\tilde m(\bx)$ are replaced by the limiting functions of estimators $\hat p(\bx)$ and $\hat m(\bx)$. This suggests that the empirical version of the minimizer above $\hat\delta(\bx)$ (whose definitions are given in (\ref{rd-learn}) and in Section \ref{sec:multi}) will be consistent with $\delta(\bx)$ if either $\hat p_1(\bx)$ or $\hat m(\bx)$ is consistent. Compared to the aforementioned robustness procedures, this estimator is robust again two types of model mis-specification, with respect to both $p_1(\bx)$ and $m(\bx)$. We call this property ``double robustness".

To compare RD-Learning and D-Learning, we compute the bias and the variance of the estimators from these two methods. Denote $\bbX = (\bbx_1, \dots, \bbx_n)^T$, $\by = (y_1, \dots, y_n)^T$, and $m(\bbX) = (m(\bx_1), \dots, m(\bx_n))^T$. Let $\bbA = \diag(a_i)$, and $\bbP_a = \diag(\hat p_{a_i}(\bx_i))$, from (\ref{rd-learn}) we derive
\begin{align}\label{betahat}
	\hat\bbeta = (\bbX^T\bbP_a^{-1}\bbX)^{-1} \bbX^T\bbA\bbP_a^{-1}(\by - \hat m(\bbX)).
\end{align}
Let $r(\bbX) = m(\bbX) - \hat m(\bbX)$ be the difference between the true main effect and its estimator. It can be shown that
\begin{align}
	\E(\hat\bbeta \mid \bbX, \bbA) &= \bbeta + (\bbX^T\bbP_a^{-1}\bbX)^{-1} \bbX^T\bbA\bbP_a^{-1}r(\bbX), \quad \label{mean_beta} \\
	\V(\hat\bbeta \mid \bbX, \bbA) &= \sigma^2 (\bbX^T\bbP_a^{-1}\bbX)^{-1} \bbX^T\bbP_a^{-2}\bbX (\bbX^T\bbP_a^{-1}\bbX)^{-1}, \quad \text{and} \nonumber \\
	\V(\hat\bbeta \mid \bbX) &= \V\left(\vphantom{\hat\bbeta} (\bbX^T\bbP_a^{-1}\bbX)^{-1} \bbX^T\bbA\bbP_a^{-1}r(\bbX) \mid \bbX\right) + \E\left(\V(\hat\bbeta \mid \bbX, \bbA) \mid \bbX\right). \label{var_beta}
\end{align}
Note that D-Learning can be viewed as a special case of RD-Learning, where $\hat m(\bx_i) = 0$ for all $i$, in which case $r(\bx_i)=m(\bx_i)$. Firstly, from (\ref{mean_beta}) we observe that the estimator by RD-Learning has a \textit{smaller bias} if $|r(\bx_i)| < |m(\bx_i)|$ holds. In the extreme that $\hat m(\bx)=m(\bx)$ which means that $r(\bx) = 0$, the RD-Learning estimator $\hat\beta$ is an unbiased estimator. Secondly, from (\ref{var_beta}) we notice that RD-Learning also has a \textit{smaller variance}. This is because in general the variability of the residual term $r(\bbX)$ is smaller than that of $m(\bbX)$, which results in a smaller value in the first term of (\ref{var_beta}). The smallest variance is achieved when $r(\bx) = 0$ (perfect estimation of $m(\bx)$.)

For high dimensional data, RD-Learning can be generalized by using a sparsity penalty. For example, we may solve a LASSO problem,
\begin{align}\label{rd-learn_lasso}
    \min_{\substack{\bbeta \in \R^p \\ \beta_0 \in \R}} \frac{1}{n} \sum_{i=1}^n \frac{1}{\hat p_{a_i}(\bx_i)} \left(y_i - \hat m(\bx_i) - a_i(\bx_i^T\bbeta + \beta_0)\right)^2 + \lambda\|\bbeta\|_1
\end{align}
with the tuning parameter $\lambda > 0$. To adopt a richer model space, we could also consider a non-linear function form for $\delta(\bx)$. For example, we may solve a kernel ridge regression problem as follows.
$$\min_{\substack{\bbeta \in \R^n \\ \beta_0 \in \R}} \frac{1}{n} \sum_{i=1}^n \frac{1}{\hat p_{a_i}(\bx_i)} \left(y_i - \hat m(\bx_i) - a_i(\bK_i^T\bbeta + \beta_0)\right)^2 + \lambda\bbeta^T\bbK\bbeta,$$
where $\bK_i$ is the $i$th column of the gram matrix $\bbK = (K(\bx_i, \bx_j))_{n\times n}$, with $K(\cdot, \cdot)$ a kernel function. Other non-linear regression models such as generalized additive model and gradient boosting can be applied in the RD-Learning framework.

\begin{figure}[!htb]
	\begin{center}
	\includegraphics[width=6.2in]{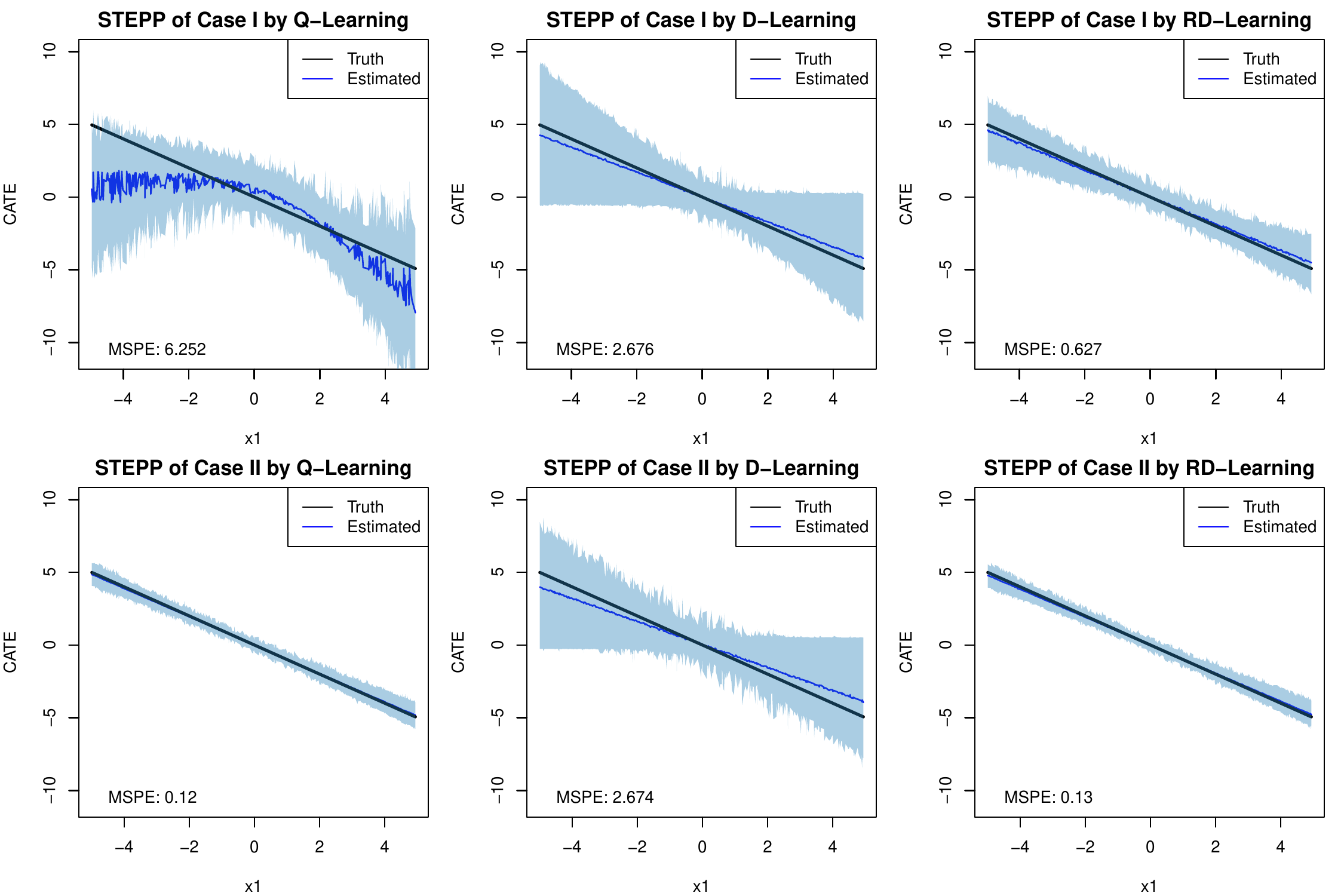}
	\end{center}
	\vspace{-1em}
	\caption{\small Subpopulation Treatment Effect Pattern Plots (STEPP) by different methods on two simulated data with $\mathcal X \subseteq \mathbb R^{20}$. Blue regions are 95\% confidence region based on 200 replications, and the black line is the true CATE. In both cases RD-Learning has a good performance.}
	\label{fig:stepp}
\end{figure}

Figure \ref{fig:stepp} compares Q-Learning, D-Learning, and RD-Learning using two toy examples. In each example, the Subpopulation Treatment Effect Pattern Plots (STEPP), a typical visualization method for exploring the heterogeneity of treatment effects \citep{bonetti2000graphical, bonetti2004patterns}, shows the relationship between the estimated CATEs and predictor $x_1$. Case I has a non-linear main effect and a linear treatment effect, where $\mu_j(\bx) = 2\cos(x_1 + \pi/4) + (3 - j)x_1/2 - \tanh(x_1)$ for $j \in \{1, -1\}$ and $p_1(\bx) = 0.2 + 0.6\ind{x_1 < 0}$. For Q-Learning, we use kernel ridge regression to estimate $\mu_j(\cdot)$; for D-Learning, we use LASSO as in \cite{qi2018d} to estimate the treatment effect $\delta(\cdot)$; in RD-Learning, we use kernel ridge regression to estimate $m(\cdot)$ and the LASSO estimator (\ref{rd-learn_lasso}) to estimate $\delta(\cdot)$. Both the main effect and the treatment effect in Case II have a linear form, with $\mu_j(\bx) = (3 - j)x_1/2 + x_2$ for $j \in \{1, -1\}$ and $p_1(\bx) = 1/2$. We use LASSO to estimate $\mu_j(\cdot)$, $m(\cdot)$, and $\delta(\cdot)$ in all three methods. Note that the true CATE is $\mu_1(\bx) - \mu_{-1}(\bx) = -x_1$ in both cases. From Figure \ref{fig:stepp}, it is clear that RD-Learning is a robust method. In particular, compared to Q-Learning, RD-Learning reduces bias in estimating CATE when the main effect tends to be mis-specified (Case I). Compared to D-Learning, RD-Learning reduces variance in both examples.

\subsection{RD-Learning in Multi-arm Case}\label{sec:multi}

In this section, we generalize RD-Learning to the case when there are more than two treatment arms. Let $A \in \mA = \{1, \dots, k\}$ be the treatment assignment. We assume the model to be
\begin{align}\label{model_multi}
    Y = m(\bX) + \delta_A(\bX) + \epsilon, \quad \text{where}\ \sum_{j=1}^k \delta_j(\bx) = 0,\ \E(\epsilon) = 0,\ \V(\epsilon | \bX) = \sigma^2(\bX) < \infty.
\end{align}
As in the binary case, $m(\cdot)$ is the main effect. $\{\delta_j(\cdot)\}_{j=1}^k$ are the treatment effects, where each of them measures the difference between the expected outcome of treatment $j$ and the main effect, \ie, $\delta_j(\bx) = \mu_j(\bx) - m(\bx)$. The sum-to-zero constraint guarantees the model is identifiable. We also allow heteroskedasticity in the white noise term $\epsilon$ to make the model more general. To estimate the treatment effect $\delta_j(\bx)$, we consider the following angle-based approach.

Angle-based approach \citep{zhang2014multicategory} is a method used in multicategory classification problem and recently it has been introduced to solve a multicategory ITR problem \citep{zhang2018multicategory, qi2019multi}. In the angle-based framework, we represent $k$ arms by $k$ vertices of a $(k-1)$-dimensional simplex, denoted by $\bW_1, \dots, \bW_k$:
$$\bW_j=\begin{cases}(k-1)^{-1/2} \bone_{k-1}, & j=1 \\
-(1 + k^{1/2}) (k - 1)^{-3/2} \bone_{k-1} + \left[k/(k-1)\right]^{1/2} \be_{j-1}, \quad & 2\leq j\leq k, \end{cases}$$
where $\bone_{k-1}$ is a $(k-1)$-dimensional vector with all elements equal to $1$ and $\be_{j-1} \in \R^{k-1}$ is a vector with the $(j-1)$th element 1 and 0 elsewhere. It is easy to check that $\|\bW_j\|=1$ and the angle $\angle(\bW_i, \bW_j)$ is the same for all $i\neq j$. The angle-based approach uses a $(k-1)$-dimensional vector-valued function $\bdf(\bx) = (f_1(\bx), \dots, f_{k-1}(\bx))^T$ as the decision function. In an ITR problem, by computing the angles between $\bdf(\bx)$ and these vertices $\bW_j$s, the optimal treatment for patient with covariates $\bx$ is chosen to be $\argmin_{j \in \mA} \angle(\bW_j, \bdf(\bx))$.

Note that for any $\bdf(\bx) \in \R^{k-1}$, the sum-to-zero constraint is satisfied for the inner products implicitly, $\sum_{j=1}^k \<{\bW_j, \bdf(\bx)} = 0$. On the other hand, for treatment effects $\{\delta_j(\bx)\}_{j=1}^k$ with $\sum_{j=1}^k \delta_j(\bx) = 0$, there is a unique $\bdf(\bx) \in \R^{k-1}$ such that $\<{\bW_j, \bdf(\bx)} = \delta_j(\bx)$ for $j \in \mA$. This motivates us to estimate the treatment effect $\delta_j(\bx)$ by $\<{\bW_j, \bdf(\bx)}$ in the angle-based framework.

\begin{theorem}\label{thm:rd_multi}
Suppose model (\ref{model_multi}) holds. Let $\tilde p_j(\bx) > 0$ be a working model for $p_j(\bx)$ and $\tilde m(\bx)$ be a working model for $m(\bx)$. Define
$$\bdf^* \in \argmin_{\bdf \in \{\mX \to \R^{k-1}\}} \E\left[\frac{1}{\tilde p_A(\bX)} (Y - \tilde m(\bX) - \<{\bW_A, \bdf(\bX)})^2\right].$$
Under Assumption \ref{asp:reg}, if either $\tilde p_j(\bx) = p_j(\bx)$ or $\tilde m(\bx) = m(\bx)$ holds for $\bx \in \mX$ almost surely and all $j \in \mA$, then $\delta_j(\bx) = \<{\bW_j, \bdf^*(\bx)}$ except on a set of measure zero.
\end{theorem}

By Theorem \ref{thm:rd_multi}, we propose the angle-based RD-Learning \footnote{Angle-based D-Learning has been studied in \cite{qi2019multi} with a different formulation.} by solving
\begin{align}\label{rd-learn_multi}
	\min_{\bdf \in \mF} \frac{1}{n} \sum_{i=1}^n \frac{1}{\hat p_{a_i}(\bx_i)} (y_i - \hat m(\bx_i) - \<{\bW_{a_i}, \bdf(\bx_i)})^2,
\end{align}
where $\mF$ is a function space. For example, we may let $\mF$ to be the linear space, \ie $\mF = \{\bdf = (f_1, \dots, f_{k-1})^T; f_j(\bx) = \bbx^T\bbeta_j, j=1, \dots, k-1\}$; or, we may consider $\mF$ in a Reproducing Kernel Hilbert Space (RKHS) with kernel function $K(\cdot, \cdot)$, \ie $\mF = \{\bdf = (f_1, \dots, f_{k-1})^T; f_j(\bx) = \sum_{i=1}^n K(\bx_i, \bx)\beta_{ij} + \beta_{0j}, j=1, \dots, k-1\}$. An $L_1$ or $L_2$ norm constraint can be also added to $\bdf$ to prevent overfitting. Denote the solution of (\ref{rd-learn_multi}) by $\hat\bdf$. Then the estimator for $j$th treatment effect is given by
\begin{align}\label{deltahat}
	\hat\delta_j(\bx) = \<{\bW_j, \hat\bdf(\bx)}.
\end{align}

Note that the binary RD-Learning introduced in Section \ref{sec:binary} is a special case of the angle-based RD-Learning. Note that when $k = 2$, we have $W_1 = 1$ and $W_2 = -1$. So $\<{W_a, f(\bx)} = f(\bx)$ for $a = 1$ and $\<{W_a, f(\bx)} = -f(\bx)$ for $a = 2$ thus (\ref{rd-learn_multi}) reduces to (\ref{rd-learn}) for the linear function space.

\subsection{Theoretical Analysis of RD-Learning}\label{sec:theory}

In this section, we study the theoretical property of $\hat\delta_j(\bx)$ defined in (\ref{deltahat}) by solving (\ref{rd-learn_multi}). Note that it suffices to consider angle-based RD-Learning since binary RD-Learning is a special case of angle-based RD-Learning. Denote $\hat\bdelta = (\hat\delta_1, \dots, \hat\delta_k)^T$. The goal of our theoretical study is to obtain the convergence rate for the prediction error (PE) of $\hat\bdelta$, defined by
$$\PE(\hat\bdelta) = \E\sum_{j=1}^k \left(\hat\delta_j(\bX) - \delta_j(\bX)\right)^2,$$
where the expectation is with respect to $\bX$. Note that since $\hat\bdelta$ depends on the training data, $\PE(\hat\bdelta)$ is a random quantity.
We consider linear models and non-linear models separately.

Before we present the main results, we make two additional assumptions for the two estimators $\hat p_j(\bx)$ and $\hat m(\bx)$.

\begin{assumption}\label{asp:p_bd}
Given estimator $\hat p_j$, we have $\|\hat p_j^{-1}(\bx) - p_j^{-1}(\bx)\|_\infty \leq r_p$ with constant $r_p > 0$.
\end{assumption}

\begin{assumption}\label{asp:m_bd}
Given estimator $\hat m$, we have $\|\hat m(\bx) - m(\bx)\|_\infty \leq r_m$ and $|Y - \hat m(\bX)| \leq C_m$ with $r_m > 0$ and $C_m > 0$.
\end{assumption}

Assumptions \ref{asp:p_bd} and \ref{asp:m_bd} state that the estimation error for $\hat p_j^{-1}(\bx)$ and $\hat m(\bx)$ are bounded with $r_p$ and $r_m$ characterizing the accuracy for both estimators. Recall that $\tilde p_j(\bx)$ and $\tilde m(\bx)$ are the limiting functions of $\hat p_j(\bx)$ and $\hat m(\bx)$ in Theorem \ref{thm:rd_multi}. The case of $\tilde p_j(\bx) = p_j(\bx)$ corresponds to $r_p \ll r_m$; the case of $\tilde m(\bx) = m(\bx)$ corresponds to $r_m \ll r_p$.

\subsubsection{Linear Function Space}

We consider a linear function space $\mF$ with a bounded $L_1$ norm:
$$\mF = \mF(p, s) \triangleq \{\bdf = (f_1, \dots, f_{k-1})^T; f_j(\bx) = \bx^T\bbeta_j + \beta_{0j}, j = 1, \dots, k-1, \sum_{j=1}^{k-1} \|\bbeta_j\|_1 \leq s\}.$$
Without loss of generality, we bound each covariate in $[-1, 1]$ for simplicity. The result still holds if we bound each covariate in $[-B, B]$ for any large number $B > 0$.

\begin{assumption}\label{asp:x_bd}
$\bX \in \mX = [-1, 1]^p$.
\end{assumption}

\begin{theorem}\label{thm:lin}
Denote $p_n$ as the dimension of the data which may depend on sample size $n$. 
Let $\mF = \mF(p_n, s_n)$ and $\tau_n = (n^{-1} \log p_n)^{1/2} \to 0$ as $n \to \infty$. Under Assumptions \ref{asp:reg} to \ref{asp:x_bd}, we have
$$\PE(\hat\bdelta) \leq \mO\left(\max\left\{(C_m + s_n)^2 \tau_n \log\tau_n^{-1}, \min\{r_1, r_2\}, d_n\right\}\right),$$
almost surely, given estimators $\hat p_j(\cdot)$ and $\hat m(\cdot)$, where $r_1 = (C_m + s_n)^2 r_p$, $r_2 = (1 + r_p) r_m^2$, and $d_n = \inf_{\bdf \in \mF(p_n, s_n)} \|\bdf - \bdf^*\|_2^2$.
\end{theorem}

\begin{remark}
Theorem \ref{thm:lin} claims that the order of $\PE(\hat\bdelta)$ is determined by three terms. The first term is the estimation error similar to the excess risk in the classification literature. As $n \to \infty$, the term will vanish for fixed $s_n$, while for fixed $n$ and $p_n$, it increases as $s_n \to \infty$ indicating a more complicated function space. The second term is determined by the accuracy of the two preliminary estimators $\hat p_j(\cdot)$ and $\hat m(\cdot)$. Specifically, $r_1$ describes the error from $\hat p_j(\bx)$ while $r_2$ describes the error from $\hat m(\bx)$. This term is small as long as either $r_p$ or $r_m$ is small, corresponding to the case when $\hat p_j(\bx)$ or $\hat m(\bx)$ is accurate. Hence, this term reflects the ``double robustness" property of the proposed estimator. The third term $d_n$ is the approximation error of the function space $\mF(p_n, s_n)$, and it will decrease as $s_n$ increases in general. The choice of $s_n$ represents a trade-off between the three terms.
\end{remark}

\begin{remark}
By Theorem \ref{thm:lin}, RD-Learning improves D-Learning in the following two aspects. Firstly, the second term in the upper bound of $\PE(\hat\bdelta)$ offers an additional way to decrease the error. Note that D-Learning is a special case of RD-Learning with $\hat m \equiv 0$, which means $r_2$ is a large number. Therefore, for D-Learning to work well, $r_1$ must be small. On the other hand, RD-Learning offers a good CATE as long as either $r_1$ or $r_2$ is small. Secondly, the estimator of RD-Learning has a smaller variance than that of D-Learning. This is because by replacing $y_i$ in D-Learning with $y_i - \hat m(\bx_i)$, the upper bound $C_m$ for $|Y - \hat m(\bX)|$ in Assumption \ref{asp:m_bd} also becomes smaller in general, which further reduces the first term in $\PE(\hat\bdelta)$. This explains the narrower confidence bands of RD-Learning in Figure \ref{fig:stepp}.
\end{remark}

Theorem \ref{thm:lin} is a general statement for the convergence rate of $\PE(\hat\bdelta)$. It neither makes assumptions on the magnitude of $r_p$ and $r_m$ which have impacts on the second term, nor assumes the true treatment effect falls in a particular function space $\mF$ which influences the third term. If we assume one of $r_p$ and $r_m$ is zero, the second term can be ignored. For example, in clinical trial $p_j(\bx)$ is known so $\hat p_j(\bx) = p_j(\bx)$ and $r_p = 0$. If we further assume $\delta_j(\bx)$ to be a linear function that only depends on finite many covariates for each $j$, the third term can be also eliminated. Since in that case, there exists a finite $p^*$ and $s^*$ such that the true population minimizer $\bdf^*$ belongs to the $\mF(p_n, s_n)$ as long as the function space we consider is large enough so that $p_n \geq p^*$ and $s_n \geq s^*$. In this case, the third term $d_n = 0$ for sufficient large $n$. The result is given in Corollary \ref{cor:lin}.

\begin{corollary}\label{cor:lin}
Let $\mF = \mF(p_n, s_n)$ and $\tau_n = (n^{-1} \log p_n)^{1/2} \to 0$ as $n \to \infty$. Suppose the true treatment effect $\delta_j(\cdot)$ depends on finite many covariates for each $j \in \mA$. Under Assumptions \ref{asp:reg} to \ref{asp:x_bd}, if either $r_p = 0$ or $r_m = 0$ holds, then we have
$$\PE(\hat\bdelta) \leq \mO(\tau_n \log\tau_n^{-1}),$$
almost surely.
\end{corollary}

From Corollary \ref{cor:lin}, we first observe that the convergence of $\PE(\hat\bdelta)$ requires that $p_n$ increases with the order at most $\exp(n)$. Secondly, since $\mO(\log x) < \mO(x^t)$ for all $t > 0$, $\PE(\hat\bdelta) \leq \mO(\tau_n^{1-t})$ for any small positive $t$. This implies that the upper bound of $\PE(\hat\bdelta)$ is almost $\mO(\tau_n) = \mO\left((n^{-1} \log p_n)^{1/2}\right)$. Furthermore, when $p_n$ is a fixed number, \ie, $p_n = \mO(1)$, the rate is almost $\mO(n^{-1/2})$. These results are coincident with most of the classical LASSO theory.

\subsubsection{Reproducing Kernel Hilbert Space}

We consider $\mF$ to be a Reproducing Kernel Hilbert Space (RKHS) to demonstrate the results for non-linear learning. The ``kernel trick" has been successfully used in many other methods like penalized regression and Support Vector Machine (SVM). There is a vast literature on RKHS. One can refer to \cite{scholkopf2001learning}, \cite{steinwart2007fast}, \cite{hofmann2008kernel}, \cite{trevor2009elements} for more details.

Let $\mH_K$ be a RKHS with kernel function $K(\cdot, \cdot)$. By the Mercer's theorem, $K$ has an eigen-expansion $K(\bx, \bx') = \sum_{i=1}^\infty \gamma_i \phi_i(\bx) \phi_i(\bx')$ with $\gamma_i \geq 0$ and $\sum_{i=1}^\infty \gamma_i^2 < \infty$. Any function in $\mH_K$ can be written as $f(\bx) = \sum_{i=1}^\infty c_i\phi_i(\bx)$ under the constraint that $\|f\|_{\mH_K}^2 = \sum_{i=1}^\infty c_i^2/\gamma_i < \infty$. Define the function space $\mF$ as
$$\mF = \mF(s) \triangleq \{\bdf = (f_1, \dots, f_{k-1})^T; f_j = f_j' + b_j, j = 1, \dots, k-1, \sum_{j=1}^{k-1} \|f_j'\|_{\mH_K}^2 \leq s^2\}.$$
Note that as in the linear case, the penalty term does not include the intercept term $b_j$. Rewrite the solution to (\ref{rd-learn_multi}) under such $\mF$ as $\hat\bdf = \hat\bdf' + \hat\bb$ where $\hat\bdf' = (\hat f_1', \dots, \hat f_k')^T$ with $f_j' \in \mH_K$. By the representer theorem \citep{wahba1990spline}, $\hat f_j'$ can be represented by
$$\hat f_j'(\bx) = \sum_{i=1}^n K(\bx_i, \bx)\hat\beta_{ij},$$
and the penalty term is written as $\|\hat f_j'\|_{\mH_K}^2 = \sum_{i=1}^n \sum_{l=1}^n K(\bx_i, \bx_l)\hat\beta_{ij}\hat\beta_{lj}$.

When developing RKHS theory, the following assumption is usually made.

\begin{assumption}\label{asp:k_bd}
The RKHS $\mH_K$ is separable and $\sup_{\bx} K(\bx, \bx) = B < \infty$.
\end{assumption}

The separability of the RKHS is commonly assumed in many papers concerning RKHS. A bounded kernel ensures that the rate of $\PE(\hat\bdelta)$ does not explode. It naturally holds for some popular kernels like Gaussian radial basis kernel, where $B = 1$. In general, it requires that $\mX$ can be covered by a compact set.

\begin{theorem}\label{thm:ker}
Let $\mF = \mF(s_n)$. Under Assumptions \ref{asp:reg}, \ref{asp:p_bd}, \ref{asp:m_bd}, and \ref{asp:k_bd}, we have
$$\PE(\hat\bdelta) \leq \mO\left(\max\left\{(C_m + Bs_n)^2 n^{-1/2} \log n, \min\{r_1, r_2\}, d_n\right\}\right),$$
almost surely, given estimators $\hat p_j(\cdot)$ and $\hat m(\cdot)$, where $r_1 = (C_m + Bs_n)^2 r_p$, $r_2 = (1 + r_p) r_m^2$, and $d_n = \inf_{\bdf \in \mF(s_n)} \|\bdf - \bdf^*\|_2^2$.
\end{theorem}

\begin{remark}
Similar to Theorem \ref{thm:lin}, there is a trade-off between the estimation error, the approximation error, and the error from $\hat p_j$ and $\hat m$ for kernel learning. $s_n$ is the tuning parameter to balance these three terms. The result also shows that compared to D-Learning, RD-Learning still enjoys a better convergence rate through a smaller $r_m$ and $C_m$.
\end{remark}

Theorem \ref{thm:ker} can be simplified in some special cases. Firstly, the second term can be ignored when $r_p$ or $r_m$ is negligible (for example, in clinical trails). Secondly, by assuming the approximation error $d_n \leq \mO(s_n^{-q})$ for some $q > 0$, which is standard in the literature on RKHS \citep{smale2003estimating}, we have a neat convergence rate by appropriately choosing $s_n$, shown in Corollary \ref{cor:ker}.

\begin{corollary}\label{cor:ker}
Let $\mF = \mF(s_n)$. Suppose $d_n = \inf_{\bdf \in \mF(s_n)} \|\bdf - \bdf^*\|_2^2 \leq \mO(s_n^{-q})$ for some $q > 0$. Under Assumption \ref{asp:reg}, \ref{asp:p_bd}, \ref{asp:m_bd}, and \ref{asp:k_bd}, if either $r_p = 0$ or $r_m = 0$ holds, then by choosing $s_n = \mO\left((n^{1/2} \log^{-1}n)^{\frac{1}{q + 2}}\right)$, we have
$$\PE(\hat\bdelta) \leq \mO\left(n^{-\frac{q}{2q + 5}}\right),$$
almost surely.
\end{corollary}

According to Corollary \ref{cor:ker}, the convergence rate of $\PE(\hat\bdelta)$ approaches to $\mO\left(n^{-1/2}\right)$ for sufficiently large $q$, corresponding to the case that $\bdf^*$ can be well approximated by a function in $\mF(s_n)$. This result is consistent with most of the learning theories under the kernel setting.

\section{Statistical Inference with Known Propensity Score}\label{sec:inference}

In this section, we consider the case when the propensity score $p_j(\bx)$ is known for each $j \in \mA$. A typical example is clinical trials, where the treatments are assigned to patients randomly with a fixed probability given $\bx$. In this case, we propose a consistent estimator for the main effect $m(\bx)$, which, according to the theories in Section \ref{sec:theory}, helps to reduce the variance of the treatment effect estimator $\hat\delta_j(\bx)$. A more fundamental contribution we make in this setting is an unbiased estimator for the treatment effect $\delta_j(\bx)$ which is allowed by the known propensity score. We also derive its asymptotic normality which is useful for constructing confidence intervals.

\subsection{A Direct Method in Estimating the Main Effect}\label{sec:me}

The RD-Learning framework we proposed in Section \ref{sec:rd-learn} is a two-step procedure. While the discussion so far focuses on the second step, \ie, estimating the treatment effect $\delta_j(\bx)$, we notice that the first step, \ie, estimating the main effect $m(\bx)$, is also important (note that there is no need to estimate $p_j(\bx)$ in this case since we assume the propensity score is known). The theoretical studies in Section \ref{sec:theory} show that an accurate $\hat m(\bx)$ reduces the variance of $\hat\delta_j(\bx)$. Moreover, the estimation of the main effect has its own value. For example, in biomedical studies, it can help researchers to identify prognostic biomarkers \citep{kosorok2019precision}.

A common method to estimate the main effect is Q-Learning. One first estimates each $\mu_j(\bx)$ for $j \in \mA$ separately, and then estimates $m(\bx)$ by taking their average. However, this main effect estimator may be inconsistent if $\mu_j(\bx)$ is mis-specified for some $j$. In addition, since the estimation of each $\mu_j(\bx)$ depends on only a portion of the data, \ie, $\{(\bx_i, a_i, y_i); a_i = j\}$, the estimator may suffer a large variance when there are very few observations in some treatment arms.
%Another method is to consider a full regression and estimate $m(\cdot)$ and $\delta_j(\cdot)$ in one step. However, the estimated the main effect depends on the accuracy of $\hat\delta_j(\cdot)$. If the function space of $\delta_j(\cdot)$ is mis-specified, $\hat m(\cdot)$ may still be inaccurate.

We propose to estimate the main effect using all the data points at the same time using weighted least square. This estimator is motivated by the important observation that under model (\ref{model_multi}),
$$\E\left[\frac{1}{p_A(\bx)} (Y - g(\bx))^2\, \Big|\, \bX = \bx\right] = \E\left[\sum_{j=1}^k (Y^*(j) - g(\bx))^2\, \Big|\, \bX = \bx\right] = \sum_{j=1}^k (\mu_j(\bx) - g(\bx))^2 + k\sigma^2(\bx),$$
and the fact that $m(\bx) = k^{-1} \sum_{j=1}^k \mu_j(\bx) = \argmin_{g(\bx)\in\R} \sum_{j=1}^k (\mu_j(\bx) - g(\bx))^2$.

Based on these observations, we propose to estimate the main effect using
\begin{align}\label{rd-learn_main}
    \hat m = \argmin_{g \in \mG} \sum_{i=1}^n \frac{1}{p_{a_i}(\bx_i)} (y_i - g(\bx_i))^2,
\end{align}
where $\mG$ is an appropriate function space. Theorem \ref{thm:rd_main} below implies that this estimator is consistent if $m\in\mG$.
\begin{theorem}\label{thm:rd_main}
Suppose the model (\ref{model_multi}) holds. Under Assumption \ref{asp:reg},
$$m \in \argmin_{g \in \{\mX \to \R\}} \E\left[\frac{1}{p_A(\bX)} (Y - g(\bX))^2\right].$$
\end{theorem}
Compared to the Q-Learning based method, the proposed method uses all the data to fit a single estimator. Besides, this estimated adopts the form of weighted least square, which can be easily generalized to and solved by many existing regression methods, such as LASSO, (kernel) ridge regression, generalized additive model, gradient boosting, and so on.

\subsection{Unbiased Treatment Effect Estimator}\label{sec:te}

In this section we focus on statistical inference of treatment effects by RD-Learning, under the special case that $\hat p_j(\bx)$ in the RD-Learning estimator is replaced the known propensity score $p_j(\bx)$ for each $j \in \mA$.

We start from the binary case under model (\ref{model_binary}). By assuming $\delta(\bx) = \bbx^T\bbeta$, the estimator $\hat\bbeta$ by RD-Learning is given by (\ref{betahat}). However, we have to point out that this estimator is biased unless $p_1(\bx_i) = 1/2$ or $\hat m(\bx_i) = m(\bx_i)$ for each $i$. In fact, according to (\ref{mean_beta}), the bias term can be explicitly written as
$$\bias(\hat\bbeta) = \E\big(\hat\bbeta\big) - \bbeta = \E\big((\bbX^T\bbP_a^{-1}\bbX)^{-1} \bbX^T\bbA\bbP_a^{-1}r(\bbX)\big).$$

\begin{remark}
To see why $\hat\bbeta$ is biased, consider a simple case where $n = 3$, $\bbX = \bone_3$, and $p_1(\bx)= p_1 \triangleq 2/3 $. Suppose we have an estimator $\hat m(\bx)$ with the residual term $r(\bx) = m(\bx) - \hat m(\bx) = 1$. It can be checked that $\bias(\hat\bbeta) = \E\left((\sum_{i=1}^3 p_{A_i}^{-1})^{-1} \sum_{i=1}^3 A_i/p_{A_i}\right)$, where the expectation is taken with respect to $\{A_i\}$. Note that there are 8 possible assignments for $\{A_i\}$: $(1, 1, 1), (1, 1, -1), \dots, (-1, -1, -1)$, with probabilities $(2/3)^3, (2/3)^2 (1/3), \dots, (1/3)^3$. So by computing this expectation explicitly, we have $\bias\big(\hat\bbeta\big) = 17/135 \neq 0$ in this toy example.
\end{remark}

To remove the bias completely, we modify (\ref{betahat}) as
\begin{align}\label{betahat_modify}
	\tilde\bbeta = \frac{1}{2} (\bbX^T\bbX)^{-1} \bbX^T\bbA\bbP_a^{-1}(\by - \hat m(\bbX)).
\end{align}
The original RD-Learning estimator (\ref{betahat}) was based the following normal equation of (\ref{rd-learn}),
\begin{align}\label{normal}
	\bbX^T\bbP_a^{-1}\bbX \bbeta = \bbX^T\bbA\bbP_a^{-1} (\by - \hat m(\bbX)).
\end{align}
Since we assume $p_{a_i}(\bx_i)$ is known, we can verify that the expectation of the left hand side of (\ref{normal}) with respect to $\bbA$ for any $\bbeta$ is
$$\E\left(\bbX^T\bbP_a^{-1}\bbX\bbeta \mid \bbX\right) = 2\bbX^T\bbX\bbeta.$$
Then by replacing the left hand side of (\ref{normal}) by its expectation, we derive the modified estimator (\ref{betahat_modify}). One can check that the bias of the modified estimator is $\bzero$.

In the multi-arm case (\ref{model_multi}), we use the angle-based approach to estimate a $(k - 1)$-dimensional decision function $\bdf(\bx) = (\bbx^T\bbeta_1, \dots, \bbx^T\bbeta_{k-1})^T$ first and then use $\<{\bW_{j}, \bdf(\bx)}$ to estimate $\delta_j(\bx)$. Specifically, we solve
$$\{\hat\bbeta_1, \dots, \hat\bbeta_{k-1}\}\in\argmin_{\{\bbeta_j\}} \frac{1}{n} \sum_{i=1}^n \frac{1}{ p_{a_i}(\bx_i)} \left(y_i - \hat m(\bx_i) - \<{\bW_{a_i}, \bdf(\bx_i)}\right)^2.$$
Denote $\hat\bbB_{(p+1) \times (k-1)}$ as $(\hat\bbeta_1, \dots, \hat\bbeta_{k-1})$. By using the similar trick as described above for the binary case, we modify $\hat\bbB$ to be an unbiased $\tilde\bbB$. The modified estimator for the treatment effect is then $\tilde \delta_j(\bx) = \<{\bW_{j}, \bbx^T\tilde\bbB} = \bbx^T\hat\bgamma_j$ where $\hat\bgamma_j\triangleq \tilde\bbB\bW_j$. We can verify that 
\begin{align}\label{gammahat}
    \hat\bgamma_j = (\bbX^T\bbX)^{-1} \bbX^T \diag\left(\ind{a_i = j} - \frac{1}{k}\right) \bbP_a^{-1}(\by - \hat m(\bbX)).
\end{align}
 
\begin{theorem}\label{thm:asymptotic}
Let $\{(\bx_i, a_i, y_i)\}_{i=1}^n$ be a random sample with $p_j(\bx_i) > 0$ for all $i$ and $j$. Assume the model (\ref{model_multi}) holds with the true treatment effect $\delta_j(\bx) = \bbx^T\bgamma_j$ and columns of $\bbX$ are linear independent. Given an estimator for the main effect $\hat m(\bx) < \infty$, denote $\bgamma = (\bgamma_1^T, \dots, \bgamma_k^T)^T$ and its estimator $\hat\bgamma = (\hat\bgamma_1^T, \dots, \hat\bgamma_k^T)^T$ with $\hat\bgamma_j$ defined in (\ref{gammahat}). Then we have
$$\E(\hat\bgamma) = \bgamma.$$

Furthermore, let $r(\bx_i) = m(\bx_i) - \hat m(\bx_i)$. Suppose $\bx_i$, $r(\bx_i)$, $p_j^{-1}(\bx_i)$, and $\sigma^2(\bx_i)$ are uniformly bounded. Denote $\bbP_{k \times k}(\bx) = \diag\left(p_j(\bx)\right)$, $\bdelta(\bx) = (\delta_1(\bx), \dots, \delta_k(\bx))^T$, and $\bbC(k) = \bbI - k^{-1}\bbJ$, where $\bbJ$ is a $k \times k$ matrix with all elements equal to $1$. If
\begin{align*}
    \bbV &= \lim_{n\to\infty} n^{-1} \bbX^T\bbX, \\
    \bbM &= \lim_{n\to\infty} n^{-1} \sum_{i=1}^n \left(\bbC(k) \diag\left((r(\bx_i) + \delta_j(\bx_i))^2\right) \bbP^{-1}(\bx_i) \bbC(k) - \bdelta(\bx_i)\bdelta(\bx_i)^T\right) \otimes \left(\vphantom{\bbP^{-1}} \bbx_i\bbx_i^T\right), \\
    \bSigma &= \lim_{n\to\infty} n^{-1} \sum_{i=1}^n \left(\bbC(k) \sigma^2(\bx_i)\bbP^{-1}(\bx_i) \bbC(k)\right) \otimes \left(\vphantom{\bbP^{-1}} \bbx_i\bbx_i^T\right)
\end{align*}
are finite and positive definite, then
$$\sqrt n (\hat\bgamma - \bgamma) \overset{\mD}{\longrightarrow} N\left(\bzero, \left(\bbJ \otimes \bbV^{-1}\right) \left(\vphantom{\bbV^{-1}} \bbM + \bSigma\right) \left(\bbJ \otimes \bbV^{-1}\right)\right).$$
\end{theorem}

Theorem \ref{thm:asymptotic} implies $\hat\bgamma_j$ is an unbiased estimator of $\bgamma_j$, and it is $\sqrt{n}$-consistent. Moreover, its variance is determined by two matrices $\bbM$ and $\bSigma$, where $\bbM$ depends on the estimator $\hat m(\cdot)$, and $\bSigma$ on the variance of $\epsilon$. Note that in D-Learning, $r(\bx_i)$ in $\bbM$ becomes $m(\bx_i)$, which is larger than $r(\bx_i)$ in RD-Learning. This also explains why RD-Learning has a smaller variance than D-Learning.

%\begin{remark}
%In the binary case, $\bbeta = \bgamma_1 = -\bgamma_2$. From Theorem \ref{thm:asymptotic}, the variance of $\tilde\bbeta$, defined in (\ref{betahat_modify}), can be written as $\V(\tilde\bbeta) = (\bbX^T\bbX)^{-1}\bbX^T \diag(\eta(\bx_i)) \bbX(\bbX^T\bbX)^{-1}$, where
%$$\eta(\bx) = \frac{(r(\bx) + \delta(\bx))^2 + \sigma^2(\bx)}{4p_1(\bx)} + \frac{(r(\bx) - \delta(\bx))^2 + \sigma^2(\bx)}{4(1 - p_1(\bx))} - \delta^2(\bx).$$
%$$\eta(\bx) = \E\left[\left(\frac{Y - \hat m(\bX)}{p_A(\bX)}\right)^2 \mid \bX = \bx\right] - \delta^2(\bx).$$
%Observe that when $\hat m(\cdot) = m(\cdot)$, the variance is minimized at $p_1(\bx) = 1/2$, corresponding to the completely randomized design. In that case, if we further assume $\sigma^2(\bx_i) = \sigma^2$ for all $i$, we have $\V(\tilde\bbeta) = \sigma^2 (\bbX^T\bbX)^{-1}$, which is the same as the result in the classical linear regression.
%\end{remark}

Theorem \ref{thm:asymptotic} can be used for constructing the confidence interval for $\bgamma$ (or the treatment effect $\delta_j$). However, the variance of $\hat\bgamma$ involves two unknown terms $\bdelta(\bx_i)\bdelta(\bx_i)^T$ and $(r(\bx_i) + \delta_j(\bx_i))^2 + \sigma^2(\bx_i)$. We may replace them by their consistent estimators, as in the heteroskedasticity literature \citep{white1980heteroskedasticity}. Specifically, $\delta_j(\bx)$ in the first term can be estimated by $\hat\delta_j(\bx_i) = \bbx_i^T\hat\bgamma_j$. For the second term, note that
\begin{align*}
	\E\left[(Y - \hat m(\bx) - \delta_A(\bx) + \delta_j(\bx))^2 \mid \bX = \bx\right] &= \E\left[(r(\bx) + \delta_j(\bx) + \epsilon)^2 \mid \bX = \bx\right] \\
	&= (r(\bx) + \delta_j(\bx))^2 + \sigma^2(\bx).
\end{align*}
This implies that we may estimate $(r(\bx_i) + \delta_j(\bx_i))^2 + \sigma^2(\bx_i)$ by $(y_i - \hat m(\bx_i) - \hat\delta_{a_i}(\bx_i) + \hat\delta_j(\bx_i))^2$.

\section{Simulation Studies}\label{sec:simulation}

We compare the proposed method with four other popular methods in estimating the treatment effect. They are Q-Learning \citep{qian2011performance}, Robust Learning \citep[R-Learning]{shi2016robust, nie2017quasi}, causal forests \citep{wager2018estimation} and D-Learning \citep{chen2017general, qi2018d}. Note that except Q-Learning and D-Learning, all other methods are two-step procedures, where the first step involves estimating either $m(\bx)$ or $\E(Y \mid \bX = \bx)$. We fix the number of covariates to be 100, where $X_1$, $X_2$ and $X_3$ are i.i.d. from $N(0, 3)$, and $X_4, \dots, X_{100}$ are i.i.d. from $\mathrm{Uniform}(0, 1)$. For each simulation setting, we let the number of observations to be $n = 50$, 100, 150, and 200. The prediction error of $\hat\delta_1$ is reported based on a testing set of size 400.

\textbf{Case I}: It is a two-arm design, with
\begin{align*}
    \mu_1(\bx) &= 2\cos(x_1 + \pi/4) + x_1 - \tanh(x_2) \quad\text{and} \\
    \mu_{-1}(\bx) &= 2\cos(x_1 + \pi/4) + 2x_1 - \tanh(x_2).
\end{align*}
The treatment assignment depends on $\bx$. Specifically, $p_1(\bx) = 0.2 + 0.6\ind{x_1 < 0}$. Since $\mu_1(\cdot)$, $\mu_{-1}(\cdot)$ and the main effect are non-linear functions of $\bx$, we consider kernel functions in Q-Learning, and in the first step of causal forests, R-Learning, and RD-Learning. On the other hand, because the treatment effect is linear, we use linear models with an $L_1$ penalty in D-Learning as well as in the second step of R-Learning and RD-Learning.\footnote{By default, the second step of causal forest uses a non-linear regression tree.}

\textbf{Case II}: This is an example to test the robustness of the proposed method against mis-specification of the main effect. In this case, we have
\begin{align*}
    \mu_1(\bx) &= \tanh(x_1) - 4/(1 + \exp(x_2 - x_1)) + 3 \quad\text{and} \\
    \mu_{-1}(\bx) &= \tanh(x_1) + 4/(1 + \exp(x_2 - x_1)).
\end{align*}
It is a randomized design with $p_1(\bx) = 1/5$. Both the main effect and the treatment effect are non-linear. Hence we are supposed to use non-linear function spaces for all the methods. However, to test the robustness of the proposed RD-Learning method, we use linear models with an $L_1$ penalty to estimate the main effect in the first step and kernel ridge regression in the second step. For comparison purposes, we adopt the same function spaces (linear and kernel) in all the other two-step procedures, and use kernel ridge regression in the one-step Q-Learning and D-Learning.

\textbf{Case III}: This is an example to test the robustness of the proposed method against mis-specification of the propensity score. In this example,
$$\mu_1(\bx) = x_1 - x_2 + x_3 \quad\text{and } \mu_{-1}(\bx) = 2x_1 - x_2.$$
The propensity score is defined as $p_1(\bx) = 2/(2 + \exp(x_1))$. In this case, we use linear models with an $L_1$ penalty in all methods and both steps. To test the robustness of RD-Learning, we deliberately use a wrong propensity score $\hat p_1(\bx) = 1/2$ instead. For comparison, we let $\hat p_1(\bx) = 1/2$ in the other methods.

\textbf{Case IV}: This is a three-arm design, with
\begin{align*}
    \mu_1(\bx) &= (x_1^2 + x_2^2 + x_3^2)/3 + x_1 - x_2, \\
    \mu_2(\bx) &= (x_1^2 + x_2^2 + x_3^2)/3 + x_2 - x_3, \quad\text{and} \\
    \mu_3(\bx) &= (x_1^2 + x_2^2 + x_3^2)/3 + x_3 - x_1.
\end{align*}
The propensity score depends on $\bx$. Specifically,
$$\left(p_1(\bx), p_2(\bx), p_3(\bx)\right) = \begin{cases} (1/2, 1/4, 1/4), \quad\text{for } x_1 \geq x_2 \text{ and } x_1 \geq x_3 \\ (1/4, 1/2, 1/4), \quad\text{for } x_2 > x_1 \text{ and } x_2 \geq x_3 \\ (1/4, 1/4, 1/2), \quad\text{for } x_3 > x_2 \text{ and } x_3 > x_1. \end{cases}$$
This setting is similar to Case I in the sense that it has a non-linear main effect and a linear treatment effect. We use the same function space as in Case I. We do not report the results by causal forests and R-Learning because currently these two methods cannot be applied to the multi-arm case directly.

\noindent{\textbf{Estimation of the treatment effect}}

\begin{figure}[!htb]
	\begin{center}
	\includegraphics[width=6.2in]{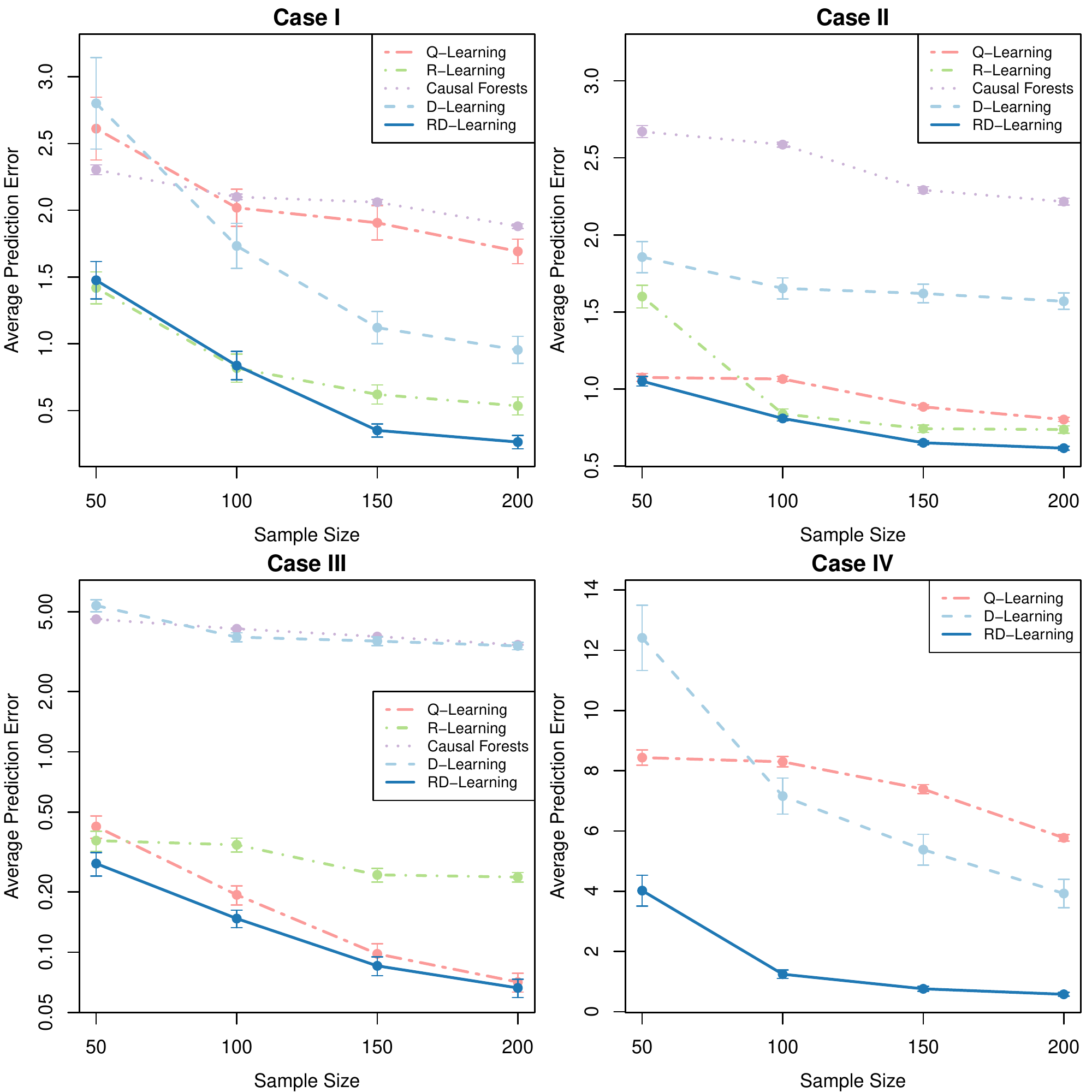}
	\end{center}
	\vspace{-1em}
	\caption{\small The average prediction error of $\hat\delta_1$ based on 200 replications with standard error by different methods. In all cases RD-Learning has the best performance.}
	\label{fig:sim_te}
\end{figure}

From Figure \ref{fig:sim_te}, we first observe that the proposed RD-Learning method has the smallest prediction error in most scenarios. Secondly, Q-Learning and D-Learning typically have a larger standard error than the two-step procedures. This is consistent with the well known intuition (see also Theorem \ref{thm:lin} and \ref{thm:ker}) that by replacing $y_i$ with $y_i - \hat m(\bx_i)$, the variance of estimators can be reduced. Thirdly, we see that RD-Learning is indeed ``doubly robust" against mis-specification of the main effect (in Case II) and the propensity score (in Case III). For the discussion below, we only focus on the three best methods, namely, R-Learning, Q-Learning, and RD-Learning. Recall that Case II is an example where we deliberately use a wrong function space for the main effect. Since R-Learning is robust against this kind of mis-specification, it has a better performance than Q-Learning. However, in Case III where we deliberately use a wrong propensity score, R-Learning has a much worse performance than Q-Learning since it relies on a correctly-specified $\hat p_j(\cdot)$. But RD-Learning is as good as, and in many cases, much better than any of these two in both settings.

\noindent{\textbf{Estimation of the main effect}}

\begin{figure}[!htb]
	\begin{center}
	\includegraphics[width=6.2in]{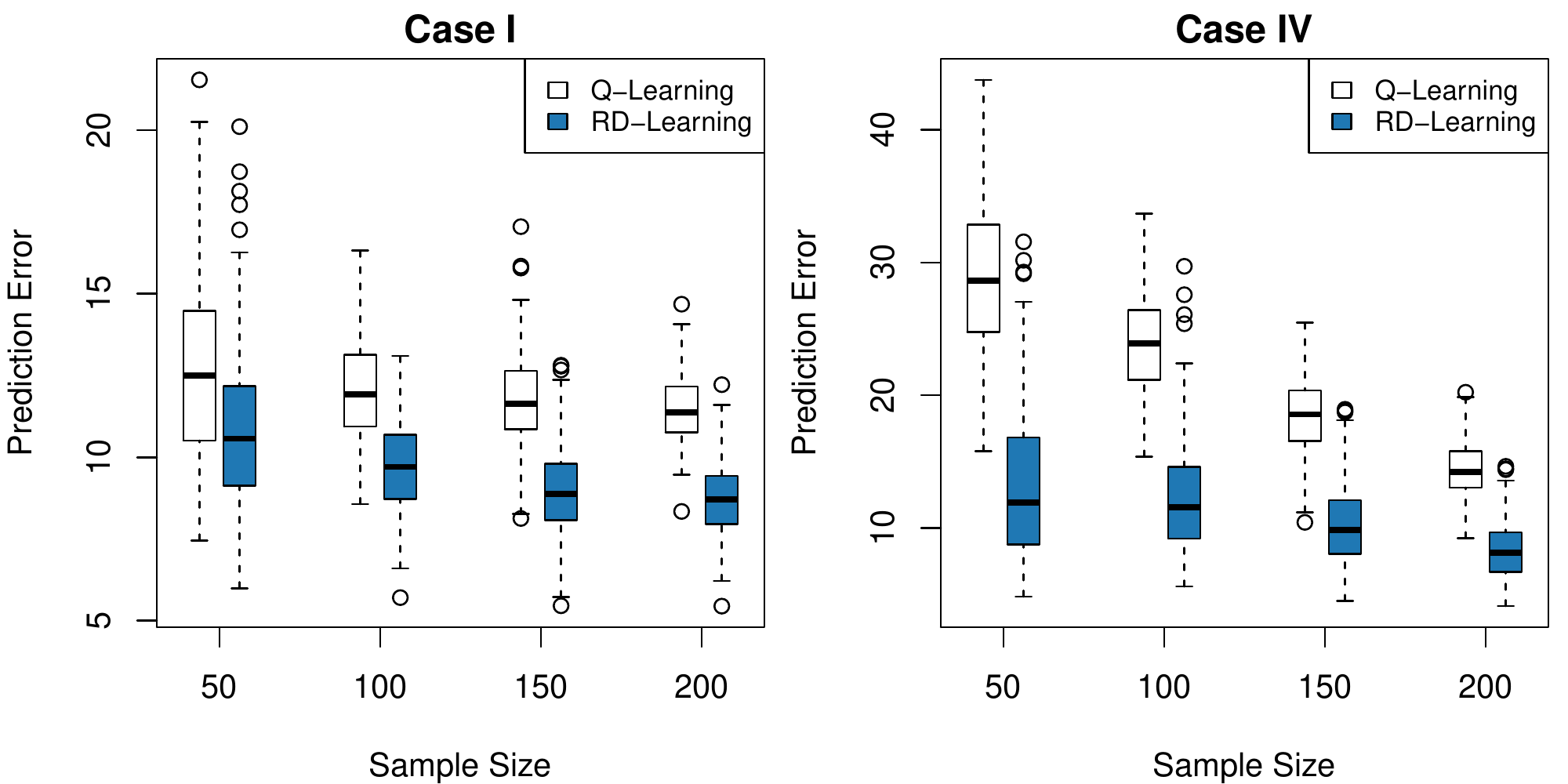}
	\end{center}
	\vspace{-1em}
	\caption{\small Boxplots for the prediction error of $\hat m$ based on 200 replications in Case I (left) and Case IV (right). The proposed method has a smaller error than Q-Learning in estimating the main effect.}
	\label{fig:sim_me}
\end{figure}

In addition to the treatment effect, we also report the estimator for the main effect using the proposed direct method in Section \ref{sec:me} and the Q-Learning method that estimates each $\mu_j$ and takes the average. Figure \ref{fig:sim_me} shows the result based on the same simulation data in Case I and Case IV. We observe that by using all the data at the same time and using propensity score as the weight, the proposed method has a better performance compared to the Q-Learning method.

\noindent{\textbf{Confidence Interval for the Coefficients}}

\begin{figure}[!htb]
	\begin{center}
	\includegraphics[width=6.2in]{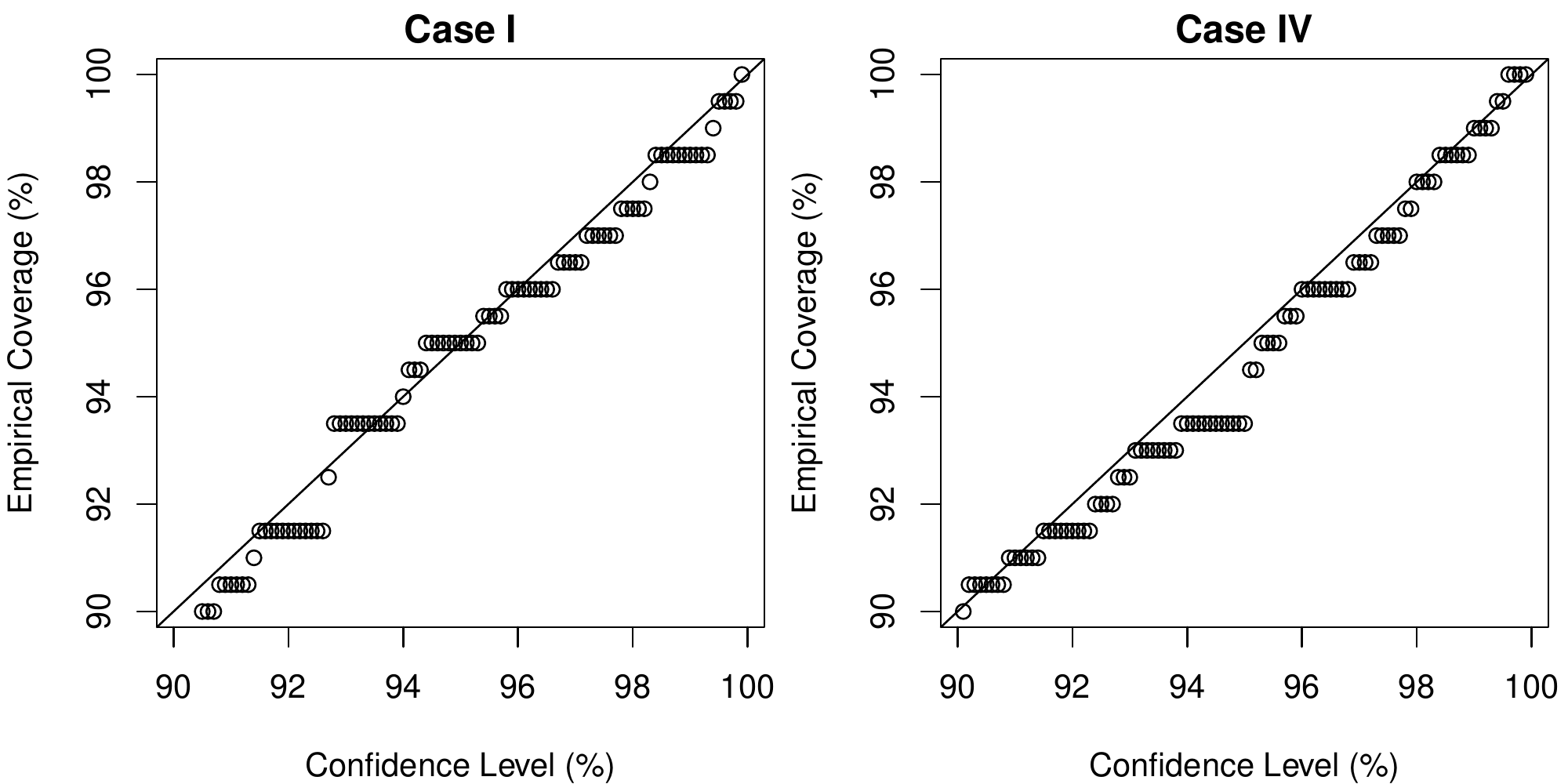}
	\end{center}
	\vspace{-1em}
	\caption{\small Nominal confidence levels ranging from 90.1\% to 99.9\% and the empirical coverage rate based on 200 replications in Case I (left) and Case IV (right) with sample size $n = 200$. The 45$^\circ$ straight line represents the ideal situation.}
	\label{fig:sim_ci}
\end{figure}

Finally, we compute the unbiased estimator defined by (\ref{gammahat}) using the data in Case I and Case IV and construct the confidence interval for the coefficient of an important covariate $x_1$. The relation between the nominal confidence level and the empirical coverage rate, defined as the proportion of the confidence intervals that cover the true parameter, is shown in Figure \ref{fig:sim_ci}. Most of the empirical coverage rates are close to the nominal confidence levels, supporting the asymptotic distribution derived in Theorem \ref{thm:asymptotic}. In the worse case scenario, for nominal level 95\%, the resulting confidence interval missed it by 1.5\% (in Case IV).

\section{Real Data Analysis}\label{sec:real}

In this section we apply RD-Learning on a real dataset from the AIDS Clinical Trials Group Study 175 \citep[ACTG175]{hammer1996trial}. The dataset includes 2,139 HIV-1 infected subjects. They were randomly assigned with equal probabilities to one of the four treatments: zidovudine (ZDV) only, ZDV with didanosine (ddI), ZDV with zalcitabine (ddC), and ddI only. The endpoint (outcome) we consider is the change of the CD4 cell count (per cubic millimeter) at $20 \pm 5$ weeks from the baseline. Note that a decrease in the number of CD4 cell count usually implies a progression to AIDS. In other words, a larger value indicates a better outcome.

To apply the proposed RD-Learning method, we first estimate the main effect using the direct estimator proposed in Section \ref{sec:me} based on the 18 variables that were measured prior to the initiation of the study. Specifically, we use the generalized additive model (GAM) to solve the weighted least square problem (\ref{rd-learn_main}). The best GAM model is selected through stepwise AIC.

For the second step in which the treatment effect is estimated, we follow the analysis of \cite{fan2017concordance} and \cite{qi2019multi} and consider only 12 variables measured at baseline as the covariates for each subject. Five of 12 covariates are continuous: age (years), weight (kilogram), Karnofsky score (on a scale of 0-100), CD4 cell counts (per cubic millimeter), and CD8 cell counts (per cubic millimeter). The rest seven are binary: hemophilia (0=no, 1=yes), homosexual activity (0=no, 1=yes), history of intravenous drug use (0=no, 1=yes), race (0=white, 1=non-white), gender (0=female, 1=male), antiretroviral history (0=naive, 1=experienced), and symptomatic indicator (0=asymptomatic, 1=symptomatic).

\begin{figure}[!htb]
	\begin{center}
	\includegraphics[width=3.6in]{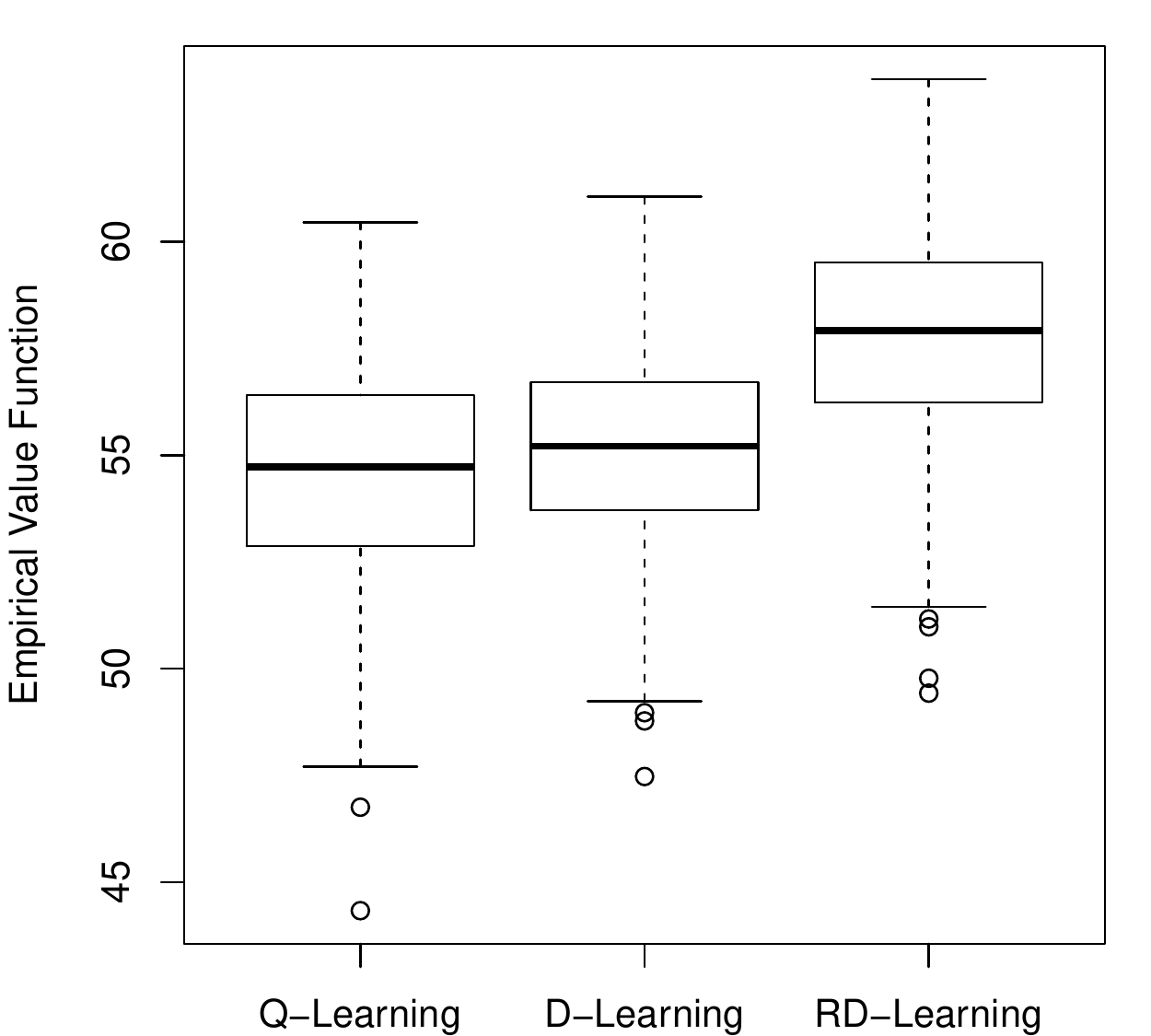}
	\end{center}
	\vspace{-1em}
	\caption{\small 5-fold cross validation scores of $V(\hat d)$ based on 400 replications by different methods for ACTG175 data. RD-Learning has the highest empirical value on average.}
	\label{fig:real_cv}
\end{figure}

We compare the performance of RD-Learning with Q-Learning and D-Learning through 5-fold cross validation. However, since it is a real data set in which the true treatment effect is not observed, the prediction error cannot be calculated. Instead of evaluating the prediction error, we first derive the estimated optimal ITR of each method by $\hat d(\bx_i) = \argmax_j \hat\delta_j(\bx_i)$. Then we calculate the empirical expected outcome under the obtained ITR $\hat d$, defined as
$$V(\hat d) = \frac{\sum_{i=1}^n \left(y_i \ind{a_i = \hat d(\bx_i)} /p_{a_i}(\bx_i)\right)}{\sum_{i=1}^n \left(\ind{a_i = \hat d(\bx_i)}/p_{a_i}(\bx_i)\right)}$$
\citep{murphy2001marginal, zhao2012estimating}. Note that in this application $V(\hat d)$ measures the average increase in CD4 cell counts (per cubic millimeter) by taking the recommended treatment. Larger value $V(\hat d)$ is preferred. Finally, we replicate the procedure for 400 times and the boxplot of $V(\hat d)$ is shown in Figure \ref{fig:real_cv}.

From Figure \ref{fig:real_cv}, we observe that RD-Learning yields the largest value, and $V(\hat d)$ of D-Learning is slightly higher than that of Q-Learning. This implies that patients would benefit more by following the recommended treatment that is based on the treatment effect estimated by RD-Learning.

To identify important biomarkers, we estimate the coefficients of the 12 covariates by (\ref{gammahat}) and compute their standard errors. The significant level of each variable using Q-Learning, D-Learning, and RD-Learning is marked in Table \ref{tab:real}.\footnote{Q-Learning is a linear regression based method with standard significance score. Since D-Learning can be viewed as a special case of RD-Learning, we derive the significance level using our method in Section \ref{sec:te}.}

\begin{table}[htb]
    \centering
    \caption{\small Significant coefficients of each treatment effect for ACTG175 Data. Each column stands for a treatment arm, and each row corresponds to a covariate. Significant coefficients and levels identified by each method are marked.}
    \label{tab:real}
    \begin{threeparttable}
    \begin{tabular}{ccccc}
    \hline\hline
    & ZDV & ZDV+ddI & ZDV+ddC & ddI \\
    \hline
    Age & & Q***, D**, RD*** & Q**, D*, RD** & \\
    Weight & & & & \\
    Hemophilia & & & & \\
    Homosexual & & Q*, D*, RD* & & Q*, D, RD* \\
    Drug use & & & Q & D, RD \\
    Karnofsky & & D**, RD & & RD \\
    Race & & Q*, D*, RD* & & \\
    Gender & & & & \\
    Antiretroviral & & & & \\
    Symptomatic & & & & \\
    CD4 Baseline & & Q**, D***, RD* & & Q**, D*, RD** \\
    CD8 Baseline & & & & \\
    \hline\hline
    \end{tabular}
    \begin{tablenotes}
    \footnotesize
    \item[*] ``Q", ``D", and ``RD" stand for Q-Learning, D-Learning, and RD-Learning, respectively.
    \item[*] Significant code example: ``Q" for p-value $< 0.1$ using Q-Learning. Similarly, ``Q*" for p-value $< 0.05$, ``Q**" for p-value $< 0.01$, and ``Q***" for p-value $< 0.001$.
    \end{tablenotes}
    \end{threeparttable}
\end{table}

\begin{figure}[!htb]
	\begin{center}
	\includegraphics[width=3.6in]{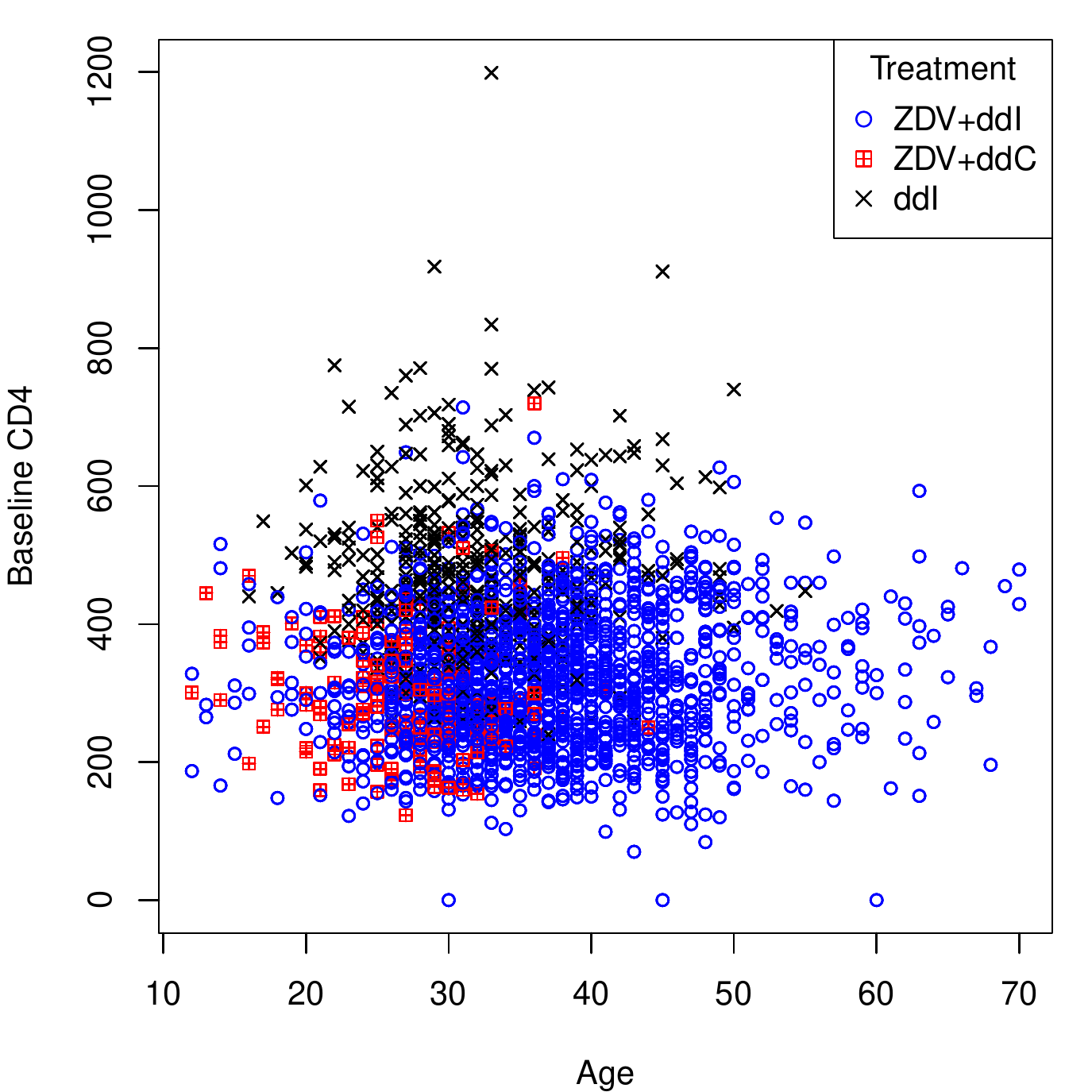}
	\end{center}
	\vspace{-1em}
	\caption{\small ACTG175 data projected on ``age" and ``CD4 baseline", with the best treatment based on the estimated treatment effect by the RD-Learning marked by different colors and symbols.}
	\label{fig:real_itr}
\end{figure}

From Table \ref{tab:real}, we observe that these three methods give similar results in general. The different patterns of those significant coefficients across different treatment effects suggest that heterogeneity does exist in these four treatment arms. For example, if we project the data on two important biomarkers ``age" and ``CD4 baseline" and mark each point according to its optimal treatment assignment estimated by RD-Learning, we can visualize how the treatment effects depend on these two biomarkers. In Figure \ref{fig:real_itr}, we first notice that the treatment ZDV is inferior to the other three treatments. This result is consistent with previous findings \citep{hammer1996trial, fan2017concordance, qi2019multi}. Furthermore, for the majority of the patients, ZDV with ddI is the best treatment. ZDV with ddC is most effective on young patients (age $< 25$), and ddI alone is better than the others for patients who have more CD4 cells (CD4 counts $> 500$ per cubic millimeter) at baseline.

\section{Conclusion}\label{sec:conclusion}

In this work, we propose a doubly robust method RD-Learning to estimate CATE under two-arm and multi-arm settings. The estimated CATE is consistent if either the model for the main effect or the model for the propensity score is correctly specified. The proposed framework is flexible enough that it can incorporate with existing base procedures such as LASSO, kernel ridge regression, generalized additive model, and so on. We also propose a direct estimation approach for the main effect and provide statistical inference tools for the treatment effects when the propensity scores are known.

There are a few possible future research directions based on this work. Firstly, by modifying the quadratic loss function, the framework can be extended to other types of outcome, such as binary outcome and survival outcome. Secondly, one may want to improve our two-step procedure to a one-step method based on (\ref{rd-learn_multi}), \ie, estimating $p_j(\bx)$, $m(\bx)$, and $\delta_j(\bx)$ simultaneously. Such CATE estimator would still enjoy a doubly-robust property while the convergence rate of $\PE(\hat\bdelta)$ may be different from the proposed method in the current paper. Thirdly, statistical inference based on RD-Learning can be investigated, so that in addition to doubly robust estimators, we may also have doubly robust confident regions. Finally the method can applied to dynamic treatment regime \citep{murphy2003optimal, robins2004optimal} by considering a multi-stage optimization problem, so that a sequence of treatment effects and the optimal treatment rules can be estimated robustly in a multi-stage clinical trial.

\bibliographystyle{asa}
\bibliography{reference}

\begin{thebibliography}{61}
\newcommand{\enquote}[1]{``#1''}
\expandafter\ifx\csname natexlab\endcsname\relax\def\natexlab#1{#1}\fi

\bibitem[{Athey and Imbens(2016)}]{athey2016recursive}
Athey, S. and Imbens, G. (2016), \enquote{Recursive partitioning for
  heterogeneous causal effects,} \textit{Proceedings of the National Academy of
  Sciences}, 113, 7353--7360.

\bibitem[{Bang and Robins(2005)}]{bang2005doubly}
Bang, H. and Robins, J.~M. (2005), \enquote{Doubly robust estimation in missing
  data and causal inference models,} \textit{Biometrics}, 61, 962--973.

\bibitem[{Beygelzimer and Langford(2009)}]{beygelzimer2009offset}
Beygelzimer, A. and Langford, J. (2009), \enquote{The offset tree for learning
  with partial labels,} in \textit{Proceedings of the 15th ACM SIGKDD
  international conference on Knowledge discovery and data mining}, pp.
  129--138.

\bibitem[{Bonetti and Gelber(2000)}]{bonetti2000graphical}
Bonetti, M. and Gelber, R.~D. (2000), \enquote{A graphical method to assess
  treatment--covariate interactions using the Cox model on subsets of the
  data,} \textit{Statistics in medicine}, 19, 2595--2609.

\bibitem[{Bonetti and Gelber(2004)}]{bonetti2004patterns}
--- (2004), \enquote{Patterns of treatment effects in subsets of patients in
  clinical trials,} \textit{Biostatistics}, 5, 465--481.

\bibitem[{Bottou et~al.(2013)Bottou, Peters, Qui{\~n}onero-Candela, Charles,
  Chickering, Portugaly, Ray, Simard, and Snelson}]{bottou2013counterfactual}
Bottou, L., Peters, J., Qui{\~n}onero-Candela, J., Charles, D.~X., Chickering,
  D.~M., Portugaly, E., Ray, D., Simard, P., and Snelson, E. (2013),
  \enquote{Counterfactual reasoning and learning systems: The example of
  computational advertising,} \textit{The Journal of Machine Learning
  Research}, 14, 3207--3260.

\bibitem[{Cao et~al.(2009)Cao, Tsiatis, and Davidian}]{cao2009improving}
Cao, W., Tsiatis, A.~A., and Davidian, M. (2009), \enquote{Improving efficiency
  and robustness of the doubly robust estimator for a population mean with
  incomplete data,} \textit{Biometrika}, 96, 723--734.

\bibitem[{Chen et~al.(2017)Chen, Tian, Cai, and Yu}]{chen2017general}
Chen, S., Tian, L., Cai, T., and Yu, M. (2017), \enquote{A general statistical
  framework for subgroup identification and comparative treatment scoring,}
  \textit{Biometrics}, 73, 1199--1209.

\bibitem[{Chipman et~al.(2010)Chipman, George, and McCulloch}]{chipman2010bart}
Chipman, H.~A., George, E.~I., and McCulloch, R.~E. (2010), \enquote{BART:
  Bayesian additive regression trees,} \textit{The Annals of Applied
  Statistics}, 4, 266--298.

\bibitem[{Dud{\'\i}k et~al.(2011)Dud{\'\i}k, Langford, and
  Li}]{dudik2011doubly}
Dud{\'\i}k, M., Langford, J., and Li, L. (2011), \enquote{Doubly robust policy
  evaluation and learning,} \textit{arXiv preprint arXiv:1103.4601}.

\bibitem[{Fan et~al.(2017)Fan, Lu, Song, and Zhou}]{fan2017concordance}
Fan, C., Lu, W., Song, R., and Zhou, Y. (2017), \enquote{Concordance-assisted
  learning for estimating optimal individualized treatment regimes,}
  \textit{Journal of the Royal Statistical Society: Series B (Statistical
  Methodology)}, 79, 1565--1582.

\bibitem[{Fan et~al.(2016)Fan, Imai, Liu, Ning, and Yang}]{fan2016improving}
Fan, J., Imai, K., Liu, H., Ning, Y., and Yang, X. (2016), \enquote{Improving
  covariate balancing propensity score: A doubly robust and efficient
  approach,} Tech. rep., Technical report, Princeton Univ.

\bibitem[{Hahn et~al.(2020)Hahn, Murray, and Carvalho}]{hahn2020bayesian}
Hahn, P.~R., Murray, J.~S., and Carvalho, C.~M. (2020), \enquote{Bayesian
  regression tree models for causal inference: regularization, confounding, and
  heterogeneous effects,} \textit{Bayesian Analysis}.

\bibitem[{Hammer et~al.(1996)Hammer, Katzenstein, Hughes, Gundacker, Schooley,
  Haubrich, Henry, Lederman, et~al.}]{hammer1996trial}
Hammer, S.~M., Katzenstein, D.~A., Hughes, M.~D., Gundacker, H., Schooley,
  R.~T., Haubrich, R.~H., Henry, W.~K., Lederman, M.~M., et~al. (1996),
  \enquote{A trial comparing nucleoside monotherapy with combination therapy in
  HIV-infected adults with CD4 cell counts from 200 to 500 per cubic
  millimeter,} \textit{New England Journal of Medicine}, 335, 1081--1090.

\bibitem[{Hill(2011)}]{hill2011bayesian}
Hill, J.~L. (2011), \enquote{Bayesian nonparametric modeling for causal
  inference,} \textit{Journal of Computational and Graphical Statistics}, 20,
  217--240.

\bibitem[{Hofmann et~al.(2008)Hofmann, Sch{\"o}lkopf, and
  Smola}]{hofmann2008kernel}
Hofmann, T., Sch{\"o}lkopf, B., and Smola, A.~J. (2008), \enquote{Kernel
  methods in machine learning,} \textit{The annals of statistics}, 1171--1220.

\bibitem[{Imbens and Rubin(2015)}]{imbens2015causal}
Imbens, G.~W. and Rubin, D.~B. (2015), \textit{Causal inference in statistics,
  social, and biomedical sciences}, Cambridge University Press.

\bibitem[{Johansson et~al.(2016)Johansson, Shalit, and
  Sontag}]{johansson2016learning}
Johansson, F., Shalit, U., and Sontag, D. (2016), \enquote{Learning
  representations for counterfactual inference,} in \textit{International
  conference on machine learning}, pp. 3020--3029.

\bibitem[{Kang and Schafer(2007)}]{kang2007demystifying}
Kang, J.~D. and Schafer, J.~L. (2007), \enquote{Demystifying double robustness:
  A comparison of alternative strategies for estimating a population mean from
  incomplete data,} \textit{Statistical science}, 22, 523--539.

\bibitem[{Knaus et~al.(2020)Knaus, Lechner, and
  Strittmatter}]{knaus2018machine}
Knaus, M.~C., Lechner, M., and Strittmatter, A. (2020), \enquote{{Machine
  learning estimation of heterogeneous causal effects: Empirical Monte Carlo
  evidence},} \textit{The Econometrics Journal}, utaa014.

\bibitem[{Kosorok and Laber(2019)}]{kosorok2019precision}
Kosorok, M.~R. and Laber, E.~B. (2019), \enquote{Precision medicine,}
  \textit{Annual review of statistics and its application}, 6, 263--286.

\bibitem[{K{\"u}nzel et~al.(2019)K{\"u}nzel, Sekhon, Bickel, and
  Yu}]{kunzel2019metalearners}
K{\"u}nzel, S.~R., Sekhon, J.~S., Bickel, P.~J., and Yu, B. (2019),
  \enquote{Metalearners for estimating heterogeneous treatment effects using
  machine learning,} \textit{Proceedings of the national academy of sciences},
  116, 4156--4165.

\bibitem[{Lu et~al.(2013)Lu, Zhang, and Zeng}]{lu2013variable}
Lu, W., Zhang, H.~H., and Zeng, D. (2013), \enquote{Variable selection for
  optimal treatment decision,} \textit{Statistical methods in medical
  research}, 22, 493--504.

\bibitem[{Moodie et~al.(2014)Moodie, Dean, and Sun}]{moodie2014q}
Moodie, E.~E., Dean, N., and Sun, Y.~R. (2014), \enquote{Q-learning: Flexible
  learning about useful utilities,} \textit{Statistics in Biosciences}, 6,
  223--243.

\bibitem[{Murphy(2003)}]{murphy2003optimal}
Murphy, S.~A. (2003), \enquote{Optimal dynamic treatment regimes,}
  \textit{Journal of the Royal Statistical Society: Series B (Statistical
  Methodology)}, 65, 331--355.

\bibitem[{Murphy(2005)}]{murphy2005generalization}
--- (2005), \enquote{A generalization error for Q-learning,} \textit{Journal of
  Machine Learning Research}, 6, 1073--1097.

\bibitem[{Murphy et~al.(2001)Murphy, van~der Laan, Robins, and
  Group}]{murphy2001marginal}
Murphy, S.~A., van~der Laan, M.~J., Robins, J.~M., and Group, C. P. P.~R.
  (2001), \enquote{Marginal mean models for dynamic regimes,} \textit{Journal
  of the American Statistical Association}, 96, 1410--1423.

\bibitem[{Nie and Wager(2017)}]{nie2017quasi}
Nie, X. and Wager, S. (2017), \enquote{Quasi-oracle estimation of heterogeneous
  treatment effects,} \textit{arXiv preprint arXiv:1712.04912}.

\bibitem[{Powers et~al.(2018)Powers, Qian, Jung, Schuler, Shah, Hastie, and
  Tibshirani}]{powers2018some}
Powers, S., Qian, J., Jung, K., Schuler, A., Shah, N.~H., Hastie, T., and
  Tibshirani, R. (2018), \enquote{Some methods for heterogeneous treatment
  effect estimation in high dimensions,} \textit{Statistics in medicine}, 37,
  1767--1787.

\bibitem[{Qi et~al.(2019)Qi, Liu, Fu, and Liu}]{qi2019multi}
Qi, Z., Liu, D., Fu, H., and Liu, Y. (2019), \enquote{Multi-Armed Angle-Based
  Direct Learning for Estimating Optimal Individualized Treatment Rules With
  Various Outcomes,} \textit{Journal of the American Statistical Association},
  1--33.

\bibitem[{Qi and Liu(2018)}]{qi2018d}
Qi, Z. and Liu, Y. (2018), \enquote{D-learning to estimate optimal individual
  treatment rules,} \textit{Electronic Journal of Statistics}, 12, 3601--3638.

\bibitem[{Qian and Murphy(2011)}]{qian2011performance}
Qian, M. and Murphy, S.~A. (2011), \enquote{Performance guarantees for
  individualized treatment rules,} \textit{Annals of statistics}, 39, 1180.

\bibitem[{Robins(2004)}]{robins2004optimal}
Robins, J.~M. (2004), \enquote{Optimal structural nested models for optimal
  sequential decisions,} in \textit{Proceedings of the second seattle Symposium
  in Biostatistics}, Springer, pp. 189--326.

\bibitem[{Robins et~al.(1994)Robins, Rotnitzky, and
  Zhao}]{robins1994estimation}
Robins, J.~M., Rotnitzky, A., and Zhao, L.~P. (1994), \enquote{Estimation of
  regression coefficients when some regressors are not always observed,}
  \textit{Journal of the American statistical Association}, 89, 846--866.

\bibitem[{Robinson(1988)}]{robinson1988root}
Robinson, P.~M. (1988), \enquote{Root-N-consistent semiparametric regression,}
  \textit{Econometrica: Journal of the Econometric Society}, 931--954.

\bibitem[{Royston and Sauerbrei(2008)}]{royston2008interactions}
Royston, P. and Sauerbrei, W. (2008), \enquote{Interactions between treatment
  and continuous covariates: a step toward individualizing therapy,} .

\bibitem[{Rubin(1974)}]{rubin1974estimating}
Rubin, D.~B. (1974), \enquote{Estimating causal effects of treatments in
  randomized and nonrandomized studies.} \textit{Journal of educational
  Psychology}, 66, 688.

\bibitem[{Scholkopf and Smola(2001)}]{scholkopf2001learning}
Scholkopf, B. and Smola, A.~J. (2001), \textit{Learning with kernels: support
  vector machines, regularization, optimization, and beyond}, MIT press.

\bibitem[{Schulte et~al.(2014)Schulte, Tsiatis, Laber, and
  Davidian}]{schulte2014q}
Schulte, P.~J., Tsiatis, A.~A., Laber, E.~B., and Davidian, M. (2014),
  \enquote{Q-and A-learning methods for estimating optimal dynamic treatment
  regimes,} \textit{Statistical science: a review journal of the Institute of
  Mathematical Statistics}, 29, 640.

\bibitem[{Shi et~al.(2016)Shi, Song, and Lu}]{shi2016robust}
Shi, C., Song, R., and Lu, W. (2016), \enquote{Robust learning for optimal
  treatment decision with NP-dimensionality,} \textit{Electronic journal of
  statistics}, 10, 2894.

\bibitem[{Signorovitch(2007)}]{signorovitch2007identifying}
Signorovitch, J.~E. (2007), \enquote{Identifying informative biological markers
  in high-dimensional genomic data and clinical trials,} Ph.D. thesis, Harvard
  University.

\bibitem[{Smale and Zhou(2003)}]{smale2003estimating}
Smale, S. and Zhou, D.-X. (2003), \enquote{Estimating the approximation error
  in learning theory,} \textit{Analysis and Applications}, 1, 17--41.

\bibitem[{Steinwart and Scovel(2007)}]{steinwart2007fast}
Steinwart, I. and Scovel, C. (2007), \enquote{Fast rates for support vector
  machines using Gaussian kernels,} \textit{The Annals of Statistics}, 35,
  575--607.

\bibitem[{Su et~al.(2009)Su, Tsai, Wang, Nickerson, and Li}]{su2009subgroup}
Su, X., Tsai, C.-L., Wang, H., Nickerson, D.~M., and Li, B. (2009),
  \enquote{Subgroup analysis via recursive partitioning.} \textit{Journal of
  Machine Learning Research}, 10.

\bibitem[{Taddy et~al.(2016)Taddy, Gardner, Chen, and
  Draper}]{taddy2016nonparametric}
Taddy, M., Gardner, M., Chen, L., and Draper, D. (2016), \enquote{A
  nonparametric bayesian analysis of heterogenous treatment effects in digital
  experimentation,} \textit{Journal of Business \& Economic Statistics}, 34,
  661--672.

\bibitem[{Tian et~al.(2014)Tian, Alizadeh, Gentles, and
  Tibshirani}]{tian2014simple}
Tian, L., Alizadeh, A.~A., Gentles, A.~J., and Tibshirani, R. (2014),
  \enquote{A simple method for estimating interactions between a treatment and
  a large number of covariates,} \textit{Journal of the American Statistical
  Association}, 109, 1517--1532.

\bibitem[{Trevor et~al.(2009)Trevor, Robert, and JH}]{trevor2009elements}
Trevor, H., Robert, T., and JH, F. (2009), \enquote{The elements of statistical
  learning: data mining, inference, and prediction,} .

\bibitem[{Turney and Wildeman(2015)}]{turney2015detrimental}
Turney, K. and Wildeman, C. (2015), \enquote{Detrimental for some?
  Heterogeneous effects of maternal incarceration on child wellbeing,}
  \textit{Criminology \& Public Policy}, 14, 125--156.

\bibitem[{Wager and Athey(2018)}]{wager2018estimation}
Wager, S. and Athey, S. (2018), \enquote{Estimation and inference of
  heterogeneous treatment effects using random forests,} \textit{Journal of the
  American Statistical Association}, 113, 1228--1242.

\bibitem[{Wahba(1990)}]{wahba1990spline}
Wahba, G. (1990), \textit{Spline models for observational data}, vol.~59, Siam.

\bibitem[{Watkins and Dayan(1992)}]{watkins1992q}
Watkins, C.~J. and Dayan, P. (1992), \enquote{Q-learning,} \textit{Machine
  learning}, 8, 279--292.

\bibitem[{Weisberg and Pontes(2015)}]{weisberg2015post}
Weisberg, H.~I. and Pontes, V.~P. (2015), \enquote{Post hoc subgroups in
  clinical trials: Anathema or analytics?} \textit{Clinical trials}, 12,
  357--364.

\bibitem[{White et~al.(1980)}]{white1980heteroskedasticity}
White, H. et~al. (1980), \enquote{A heteroskedasticity-consistent covariance
  matrix estimator and a direct test for heteroskedasticity,}
  \textit{econometrica}, 48, 817--838.

\bibitem[{Zhang et~al.(2012)Zhang, Tsiatis, Laber, and
  Davidian}]{zhang2012robust}
Zhang, B., Tsiatis, A.~A., Laber, E.~B., and Davidian, M. (2012), \enquote{A
  robust method for estimating optimal treatment regimes,} \textit{Biometrics},
  68, 1010--1018.

\bibitem[{Zhang et~al.(2013)Zhang, Tsiatis, Laber, and
  Davidian}]{zhang2013robust}
--- (2013), \enquote{Robust estimation of optimal dynamic treatment regimes for
  sequential treatment decisions,} \textit{Biometrika}, 100, 681--694.

\bibitem[{Zhang et~al.(2018)Zhang, Chen, Fu, He, Zhao, and
  Liu}]{zhang2018multicategory}
Zhang, C., Chen, J., Fu, H., He, X., Zhao, Y., and Liu, Y. (2018),
  \enquote{Multicategory Outcome Weighted Margin-based Learning for Estimating
  Individualized Treatment Rules,} \textit{Statistica Sinica}.

\bibitem[{Zhang and Liu(2014)}]{zhang2014multicategory}
Zhang, C. and Liu, Y. (2014), \enquote{Multicategory angle-based large-margin
  classification,} \textit{Biometrika}, 101, 625--640.

\bibitem[{Zhao et~al.(2012)Zhao, Zeng, Rush, and Kosorok}]{zhao2012estimating}
Zhao, Y., Zeng, D., Rush, A.~J., and Kosorok, M.~R. (2012), \enquote{Estimating
  individualized treatment rules using outcome weighted learning,}
  \textit{Journal of the American Statistical Association}, 107, 1106--1118.

\bibitem[{Zhao et~al.(2019)Zhao, Laber, Ning, Saha, and
  Sands}]{zhao2019efficient}
Zhao, Y.-Q., Laber, E.~B., Ning, Y., Saha, S., and Sands, B.~E. (2019),
  \enquote{Efficient augmentation and relaxation learning for individualized
  treatment rules using observational data.} \textit{Journal of Machine
  Learning Research}, 20, 1--23.

\bibitem[{Zhao et~al.(2014)Zhao, Zeng, Laber, Song, Yuan, and
  Kosorok}]{zhao2014doubly}
Zhao, Y.-Q., Zeng, D., Laber, E.~B., Song, R., Yuan, M., and Kosorok, M.~R.
  (2014), \enquote{Doubly robust learning for estimating individualized
  treatment with censored data,} \textit{Biometrika}, 102, 151--168.

\bibitem[{Zhou et~al.(2017)Zhou, Mayer-Hamblett, Khan, and
  Kosorok}]{zhou2017residual}
Zhou, X., Mayer-Hamblett, N., Khan, U., and Kosorok, M.~R. (2017),
  \enquote{Residual weighted learning for estimating individualized treatment
  rules,} \textit{Journal of the American Statistical Association}, 112,
  169--187.

\end{thebibliography}

\end{document}